\documentclass[10pt]{article}
\usepackage{fullpage,graphicx,subfigure,mathdots,mathpazo,color}
\usepackage{amsmath,amscd,tikz,mathrsfs}
\usepackage[normalem]{ulem}
\usepackage{amsmath}
\usepackage{setspace,booktabs}%äœ¿çšéŽè·å®å

\usepackage{epsfig,amsmath,graphicx,amssymb,overpic}

\usepackage{graphicx}% Include figure files
\usepackage{dcolumn}% Align table columns on decimal point
\usepackage{bm}% bold math
\usepackage{graphicx}
\usepackage{subfigure}
\usepackage{epsfig,amsmath,graphicx,amssymb,overpic,cite}

\usepackage{graphicx}% Include figure files
\usepackage{dcolumn}% Align table columns on decimal point
\usepackage{bm}% bold math
\usepackage{graphicx}
\usepackage{subfigure}
\usepackage{amsthm}

\newcommand{\ee}{\mathrm{e}}

\usepackage[T1]{fontenc}
\usepackage{amsmath, amsfonts, amssymb, amsthm}
\usepackage{geometry} 
\usepackage{hyperref} \hypersetup{colorlinks=true, linkcolor=blue, citecolor=red}
\usepackage{enumitem}

\usepackage{graphicx}
\usetikzlibrary{positioning,arrows.meta}

\usepackage{graphicx}      % insert graphic
\usepackage{titlesec}
\def\bee{\begin{eqnarray}}
\def\ene{\end{eqnarray}}
\def\bes{\begin{subequations}}
\def\ees{\end{subequations}}

\usepackage{color}

\def\v{\vspace{0.06in}}
\def\be{\begin{equation}}
\def\ee{\end{equation}}
\def\bee{\begin{eqnarray}}
\def\ene{\end{eqnarray}}
\def\bes{\begin{subequations}}
\def\ees{\end{subequations}}

\theoremstyle{plain}
\newtheorem{theorem}{Theorem}[section]

\theoremstyle{definition}
\newtheorem{definition}[theorem]{Definition}

\newtheorem{remark}[theorem]{Remark}

\numberwithin{equation}{section}

\newtheorem{assumption}[theorem]{Assumption}
\newtheorem{RH}[theorem]{Riemann-Hilbert Problem}

\def\be{\begin{equation}}
\def\ee{\end{equation}}
\def\bee{\begin{eqnarray}}
\def\ene{\end{eqnarray}}
\def\bes{\begin{subequations}}
\def\ees{\end{subequations}}

\def\v{\vspace{0.1in}}

\allowdisplaybreaks[4]

\numberwithin{equation}{section}

\begin{document}
%%%%%%%%%%%%%%%%%%%%%%%%%%%%%%%%%%%%%%%%%%%%%%%%%%%%%%%%%%%%%%%%%%%%%%%%%%%%%%%%%

\baselineskip=15pt \renewcommand {\thefootnote}{\dag}
\renewcommand
{\thefootnote}{\ddag} \renewcommand {\thefootnote}{ }

\pagestyle{plain}

\begin{center}
\baselineskip=18pt \leftline{} \vspace{0.05in} {\Large\bf Long-time asymptotic behavior for the defocusing Hirota equation on a finite-genus algebro-geometric background} \\[0.2in]
\end{center}

\begin{center}
{ Taohua Luo$^{1,2}$,\,\, Zhenya Yan$^{3,2,1,*}$,\,\,  Guoqiang Zhang$^{1}$}
\footnote{$^{*}${\it Email address}: zyyan@mmrc.iss.ac.cn (Corresponding author)}  \\[0.1in]
\baselineskip=11pt {\footnotesize  $^1${\it State Key Laboratory of Mathematical Sciences, Academy of Mathematics and Systems Science, \\ \footnotesize Chinese Academy of Sciences,  Beijing 100190, China}}\\
{\footnotesize  $^2${\it School of Mathematical Sciences, University of Chinese Academy of Sciences, Beijing 100049, China}}\\
{\footnotesize  $^3${\it School of Mathematics and Information Science, Zhongyuan University of Technology, Zhengzhou 450007, China}} \\
\end{center}

\v

\noindent\rule{\textwidth}{0.6pt}

\noindent {\bf Abstract}

\noindent
In this paper, we investigate the long-time asymptotics for the solution of the Cauchy problem of the defocusing Hirota equation on a finite-genus algebro-geometric background in the whole $(x,t)$-half-plane, whose method is mainly based on a Riemann-Hilbert  (RH) formulation and Deift-Zhou nonlinear steepest descent method.
The critical values of the phase function in the associated RH problem divide the space-time plane into four regions, in which the leading-order term is given by a phase-shifted finite-genus algebro-geometric solution. The subleading behavior depends on the region: the correction is of order $t^{-1/3}$ and is governed by a Painlev\'e-XXXIV model RH problem in the transition regions; the leading radiation is of order $t^{-1/2}$ in the Zakharov--Manakov region; and the error is $O(t^{-1})$ in the fast-decay region. These results can also be extended to other higher-order members of the AKNS hierarchy.
%The results extend the finite-genus asymptotic theory for the defocusing nonlinear Schrödinger equation to the defocusing Hirota equation.

\vspace{1em}
\noindent\textbf{Keywords}\,  Defocusing Hirota equation; finite-genus algebro-geometric background; Riemann-Hilbert problem;  long-time asymptotics; Painlev\'e-XXXIV equation

\vspace{1em}
\noindent\textbf{Mathematics Subject Classification}\,\, 35Q51 $\cdot$ 35Q15 $\cdot$ 37K15 $\cdot$ 35P20 $\cdot$ 35C20 $\cdot$ 35G25

\noindent\rule{\textwidth}{0.6pt}

\begin{spacing}{1.2}
\baselineskip=13pt
\tableofcontents
\vspace{-0.2in}
\end{spacing}

%%%%%%%%%%%%%%%%%%%%%%%%%%%%%%%%%%%%%%%%%%%%%%%%%%%%%%%%%%%%%%%%%%%%

\baselineskip=15pt

\section{Introduction}

\quad In this paper, we  consider the long-time asymptotics of the  solution for Cauchy problem of the defocusing Hirota equation \cite{Hirota1973}
\begin{equation}\label{Hirota}
iq_t+\alpha(q_{xx}-2|q|^2q)+i\beta(q_{xxx}-6|q|^2q_x)=0,\quad  (x,t)\in \mathbb{R}\times\mathbb{R}^+,
\end{equation}
under a finite-genus algebro-geometric background:
\begin{equation}\label{background}
q(x,0)=q_0(x)\sim q^{(AG)}(x,t)\big|_{t=0},\qquad x\to\pm\infty,
\end{equation}
where $q=q(x,t)$ is an envelope field, $\alpha,\beta\in\mathbb R$ are real parameters, $q^{(AG)}(x,t)=q^{(AG)}(x,t;\mathbf E,\widehat{\mathbf E},\boldsymbol\phi)$ denotes a finite-genus algebro-geometric solution of  \eqref{Hirota}, which can be found by
using the Baker-Akhiezer function on the hyperelliptic Riemann surface of genus $n$ defined by ~\cite{Belokolos1994,Farkas-Riemann92,Gesztesy2003,Kamvissis2003}.
\begin{equation}\label{RSurface}
w^2(z)=\prod_{j=0}^{n}(z-E_j)(z-\widehat E_j),\qquad E_0<\widehat E_0<E_1<\widehat E_1<\cdots<E_n<\widehat E_n,
\end{equation}
where $\mathbf E=(E_0,\ldots,E_n),\, \widehat{\mathbf E}=(\widehat E_0,\ldots,\widehat E_n),\, \boldsymbol\phi=(\phi_1,\ldots,\phi_n)$ are real-valued vectors
with $n\in\mathbb N_0$. For completeness of the Cauchy problem, the perturbation is assumed to be integrable relative to the algebro-geometric background,
\begin{equation}\label{qRestrict}
\int_{\mathbb R}|q(x,t)-q^{(AG)}(x,t)|\,dx<\infty,\qquad t\ge0.
\end{equation}
In the asymptotic analysis below, we impose a compact perturbation assumption on the initial data, namely $q_0(x)=q^{(AG)}(x,0)$ for $|x|$ sufficiently large. This assumption ensures good analytic properties of the scattering data and allows us to focus on the essential Riemann--Hilbert steepest descent mechanism.

 The Hirota equation (\ref{Hirota}), originally introduced by Hirota in the study of exact envelope-soliton solutions \cite{Hirota1973}, is one of the most important higher-order integrable extensions of the focusing ($\sigma=1$) and defocusing ($\sigma=-1$) nonlinear Schr\"odinger (NLS) equation \cite{Zakharov1972}
  \begin{equation}\label{NLSEquation}
iq_t+q_{xx}+2\sigma |q|^2q=0,\quad \sigma=\pm1,
\end{equation}
and extension of the focusing ($\sigma=1$) and defocusing ($\sigma=-1$) complex modified Korteweg-de Vries (cmKdV) equations \cite{Wadati1972} (Notice that Eq.~(\ref{mKdVEquation}) reduces to the mKdV equation as $q=\bar{q}$, where the bar denotes the complex conjugate):
\begin{equation}\label{mKdVEquation}
q_t+q_{xxx}+6\sigma|q|^2q_x=0, \quad \sigma=\pm1.
\end{equation}
Such higher-order models naturally occur in nonlinear optics and dispersive wave theory, where the standard NLS approximation is often refined by higher-order dispersion and nonlinear self-steepening effects \cite{Hasegawa1995}. The Hirota equation is also connected, via the Hasimoto transformation and its higher-order corrections, with vortex filament motion and axial-flow effects in incompressible fluids \cite{Hasimoto1972,Fukumoto1991}.
%Thus \eqref{Hirota} may be viewed as a linear combination of two commuting flows (i.e., second-order NLS flow and third-order cmKdV flow) in the AKNS hierarchy \cite{Ablowitz1974}.
When $\beta=0$, equation \eqref{Hirota} reduces to the defocusing NLS equation \eqref{NLSEquation} with $\sigma=-1$; when $\alpha=0$, it reduces to the defocusing cmKdV equation (\ref{mKdVEquation}) with $\sigma=-1$. This mixture of the NLS and cmKdV flows in the AKNS hierarchy \cite{Ablowitz1974} makes the Hirota equation a natural model for studying higher-order effects within the inverse-scattering framework \cite{Faddeev1987}.
Although the parameters $\alpha$ and $\beta$ in \eqref{Hirota} are allowed to be arbitrary real numbers, it is sufficient for the purpose of the present asymptotic analysis to consider the case $\alpha\ge 0,\, \beta>0$.
Indeed, the spatial reflection $q(x,t)\mapsto q(-x,t)$ changes $(\alpha,\beta)$ into $(\alpha,-\beta)$, while the complex conjugation $q(x,t)\mapsto \overline{q(x,t)}$ changes $(\alpha,\beta)$ into $(-\alpha,\beta)$. Hence every case with $\beta\ne0$ can be reduced to $\alpha\ge0$, $\beta>0$. The case $\beta=0$ corresponds to the defocusing NLS equation and has already been treated in the finite-genus algebro-geometric setting \cite{Fan2026}. Therefore, throughout the rest of this paper we focus on the genuinely Hirota case $\alpha\ge0$, $\beta>0$.
The NLS, cmKdV, and their higher-order extensions \cite{Ablowitz1991,Newell1985} occupy a central position in the theory of nonlinear wave propagation \cite{Hasegawa1995}, inverse scattering \cite{Ablowitz1981,Gardner1967}, and Riemann--Hilbert problems \cite{Deift1993,Deift1999a}.

The inverse scattering transform (IST) for AKNS-type systems \cite{Ablowitz1974} was developed in the classical works \cite{Gardner1967,Zakharov1979}, and its rigorous formulation with possible spectral singularities was further studied in \cite{Zhou1989}. Within this framework, the Riemann--Hilbert (RH) formulation of inverse scattering \cite{Beals1984} has become one of the most powerful tools for both exact solution theory \cite{Belokolos1994} and long-time asymptotic analysis \cite{Zakharov1976}.
The long-time asymptotic behaviors of solutions to integrable nonlinear wave equations is a fundamental subject in nonlinear dispersive PDEs \cite{Zakharov1976}. For rapidly decaying initial data, the IST converts the Cauchy problem into a matrix RH problem whose jump matrix contains oscillatory exponential factors depending on a large parameter $t$. The nonlinear steepest descent method of Deift and Zhou \cite{Deift1993} provides a systematic way of deforming the associated contour and reducing the original oscillatory RH problem to explicitly solvable model problems. Since its introduction, this method and its refinements have been successfully applied to the mKdV equation \cite{Deift1993}, the NLS equation \cite{Deift2003}, the KdV equation \cite{Deift1994,Deift1997} and many other integrable models \cite{Kamvissis2003,Miller2006}. A crucial ingredient in many nonlinear steepest descent analyses is the so-called $g$-function mechanism, which is designed to normalize the oscillatory phase and deform the original RH problem into a model problem that is explicitly solvable. This mechanism also provides a systematic way of constructing local parametrices near critical points and deriving rigorous error estimates \cite{Ablowitz1974,Zakharov1979,Zhou1989,Kamvissis2003,Miller2006}.

Many types of integrable structures of the Hirota equation have been investigated from several complementary viewpoints. For example, the bilinear method \cite{Hirota1973,Hirota2004} yields exact multisoliton solutions \cite{Hirota1973}, while inverse scattering and matrix triplet methods produce explicit reflectionless solutions, including solitons, breathers and multipole soliton solutions \cite{chensy2019,Demontis2015,weng21,zhang20}. The Darboux transform and its generalization were used to find rogue waves of the Hirota equation ~\cite{gu2005, tao2012}. Nonlinear stability of multi-solitons was studied for the Hirota equation \cite{xiao2023}. For the Hirota equation with decaying initial data, Huang {\it et al} derived a RH representation and applied the Deift--Zhou nonlinear steepest descent method to obtain its long-time asymptotic formulae \cite{Huang2015}. Related long-time asymptotic problems on the half-line were also considered in \cite{Guo2018}. Chen {\it et al}~\cite{chen-h21} and Pemg {\it et al}~\cite{peng26} studied the long-time asymptotics for the focusing and defocusing Hirota equations with non-zero constant boundary conditions.
More recently, Xun {\it et al} studied the transition region for the defocusing Hirota equation with Schwartz initial data, and showed that the leading asymptotics in the critical region can be expressed in terms of a Painlev\'e II transcendent \cite{Xun2022}. In parallel, the algebro-geometric integration of the Hirota equation and the corresponding Baker-Akhiezer function have also been studied by means of RH method \cite{Cao2024}. These works demonstrate that the Hirota equation admits a rich asymptotic structures, while also showing that the third-order flow introduces phase functions and critical-point configurations that are more complicated than those of the standard NLS equation.

Most of the above-mentioned asymptotic results for the Hirota equation concern either zero boundary conditions, rapidly decaying perturbations of trivial backgrounds, or non-zero constant backgrounds. However, many physically and mathematically relevant nonlinear wave patterns are not perturbations of zero, but rather perturbations of periodic, quasi-periodic or finite-gap backgrounds \cite{Dubrovin1981,Its1975}. From the point of view of finite-gap integration, algebro-geometric solutions provide the natural class of nonlinear quasi-periodic backgrounds for integrable systems. Such solutions are constructed on hyperelliptic Riemann surfaces by means of Abelian differentials, Abel maps, Riemann theta functions and Baker-Akhiezer functions; see, for example, \cite{Belokolos1994,Gesztesy2003}. In the Riemann--Hilbert approach, the finite-gap background is encoded by a model problem with constant jumps along spectral bands \cite{Kamvissis2003,BoutetdeMonvel2008,Deift1999b}. For the NLS equation, the planar unimodular Baker--Akhiezer function and its RH characterization were developed in \cite{Kotlyarov2017}.

The study of perturbations of nonzero or finite-gap backgrounds has attracted considerable attention in recent years. Even in the zero-boundary setting, the $(x,t)$-plane may contain narrow transition regions where the asymptotic behavior differs from that in the standard oscillatory or decay sectors; see, for instance, the collisionless-shock and transition-region analyses for KdV and related models \cite{Deift1994,Deift1997}. In the presence of nonzero or finite-gap backgrounds, these transition phenomena become more intricate because the associated RH problems involve spectral bands, branch points and nontrivial background parametrices. For the focusing NLS equation with decaying data, nonzero, step-like or oscillatory backgrounds, several long-time asymptotic scenarios have been analyzed by RH methods \cite{Biondini2021,Borghese2018,BoutetdeMonvel2011,BoutetdeMonvel2021}.
%Long-time asymptotics for the focusing NLS equation with nonzero boundary conditions in the presence of a discrete spectrum were studied in \cite{Biondini2021}, while the long-time asymptotic behavior of the focusing NLS equation with decaying data was analyzed in \cite{Borghese2018}.
The step-like oscillating case also exhibits rich genus-dependent structures, including a genus-three sector studied in \cite{BoutetdeMonvel2022}. Beyond the NLS setting, finite-gap background perturbations have been investigated for the periodic Toda lattice and the KdV equation in \cite{Kamvissis2012,MikikitsLeitner2012}.
%Very recently, Ma and Fan studied the focusing NLS equation with a finite-genus algebro-geometric background and showed that the nature of the local model may depend on the parity of the genus: Painlev\'e II parametrices arise in odd-genus backgrounds, whereas parabolic-cylinder parametrices appear in even-genus backgrounds \cite{Ma2026}.
For the defocusing NLS equation with a finite-genus algebro-geometric background, Fan {\it et al} recently obtained long-time asymptotics in four distinct space-time regions and showed that Painlev\'e XXXIV transcendents arise naturally in the transition regions \cite{Fan2026}. We also refer to \cite{Its2008} for the appearance of the Painlevé XXXIV transcendent in critical edge asymptotics of random matrix ensembles. These works reveal that finite-gap backgrounds lead to a delicate interaction between the topology of the spectral curve, the distribution of stationary phase points, theta-function parametrices and Painlev\'e-type local models.

The Cauchy problem of the defocusing Hirota equation (\ref{Hirota})-(\ref{background}) considered here is closely related to, but substantially different from, the defocusing NLS problem on a finite-genus algebro-geometric background \cite{Fan2026}. In comparison with the NLS equation, the Hirota equation (\ref{Hirota}) contains an additional third-order flow. This changes the phase function, modifies the critical rays, and leads to a different signature table. In the present problem, the phase function has the form
\begin{equation}\label{phasefunction}
\theta(z;\xi)=-(f(z)-f_0)\xi-\alpha(g(z)-g_0)-\beta(h(z)-h_0), \qquad \xi=\frac{x}{t},
\end{equation}
where $f(z), g(z), h(z)$ defined by (\ref{fgh})  are normalized Abelian integrals of the second kind associated with the $x$-flow, the NLS time flow and the complex mKdV time flow, respectively \cite{Belokolos1994,Dubrovin1981}, and $f_0,\, g_0,\, h_0$ are given by (\ref{fgh0}). The critical point structure of this phase function determines the space-time decomposition used in the asymptotic analysis. Compared with the NLS case, the additional Abelian integral $h(z)$ associated with the third-order AKNS flow modifies the phase geometry, the critical rays and the associated signature table, and hence requires a separate steepest descent analysis.

The main purpose of this paper is to obtain the long-time asymptotics of the solution $q(x,t)$ to the Cauchy problem \eqref{Hirota}, \eqref{background},  \eqref{qRestrict} in the whole $(x,t)$-half-plane $t>0$.  The quantities $\xi_j$ and $\widehat\xi_j$ determined by the critical point conditions divide the $(x,t)$-plane into four types of asymptotic regions: two transition regions, denoted by region I and region II, the Zakharov--Manakov \cite{Zakharov1976} region III , and the fast-decay region IV. Our results show that the leading term in all regions is given by the finite-genus algebro-geometric background with a phase shift determined by the reflection data. More precisely, the leading contribution has the form
\begin{equation}\label{LeadingTerm}
q^{lt}=e^{-2\delta(\infty)} q^{(AG)}(x,t;\mathbf E,\widehat{\mathbf E},\boldsymbol\phi-\boldsymbol\delta),
\end{equation}
where $\boldsymbol\delta$ and $\delta(\infty)$ are determined by the scalar $\delta$-function associated with the corresponding deformation of the RH problem.

The subleading terms depend essentially on the space-time region. In the two transition regions, where $\xi=x/t$ approaches a critical value at the rate $t^{-2/3}$, the correction term is of order $t^{-1/3}$ and is expressed in terms of a model RH problem associated with the Painlev\'e XXXIV equation \cite{Fokas2006}. The RH approach to Painlev\'e transcendents has been systematically developed in \cite{Fokas2006}, and Painlev\'e-type asymptotics have appeared in a variety of nonlinear wave and random matrix problems \cite{Buckingham2014,Claeys2008}. In the present setting, for $|\xi-\widehat\xi_j|t^{2/3}\le C$ or $|\xi-\xi_j|t^{2/3}\le C$, the solution admits an expansion of the form
\begin{equation}\label{transition-asymptotics}
q(x,t)=e^{-2\delta(\infty)} q^{(AG)}(x,t;\mathbf E,\widehat{\mathbf E},\boldsymbol\phi-\boldsymbol\delta) +H\,\nu(s)t^{-1/3}+O(t^{-\varepsilon}), \qquad \varepsilon\in(1/3,2/3),
\end{equation}
where $s$ is the scaled transition variable, $\nu(s)$ is determined by the Painlev\'e XXXIV model, and $H$ is an explicit coefficient depending on the local behavior of the global parametrix and the phase near the relevant branch point. In the Zakharov-Manakov region \cite{Zakharov1976}, where two simple saddle points contribute to the leading radiation, the first correction is of order $t^{-1/2}$. In the fast-decay region, there are no real stationary phase contributions to the leading-order radiation term, and the solution is approximated by the shifted algebro-geometric background with an error of order $O(t^{-1})$.

The proof is based on a nonlinear steepest descent analysis of RH problems \cite{Deift1993} adapted to a finite-genus background. We first construct Jost solutions normalized by the algebro-geometric Baker--Akhiezer function \cite{Kotlyarov2017} and formulate two equivalent RH problems. The use of two RH problems is convenient because different factorizations of the jump matrix are required in different sign regions of the phase. We then introduce scalar $\delta$-functions to remove the non-decaying diagonal part of the jump matrices and to incorporate the reflection coefficients into the phase-shifted finite-gap background. The global parametrix is built from the finite-genus Baker-Akhiezer function and can be expressed in terms of Riemann theta functions. Near the critical branch points, however, the global parametrix is no longer sufficient; one must construct local parametrices that match the global one on the boundary of small neighborhoods. In the transition regions, the local model is governed by a Painlev\'e XXXIV Riemann--Hilbert problem. Finally, a standard small-norm argument yields the error estimates and allows us to recover the asymptotic expansion of $q(x,t)$ from the large-$z$ behavior of the solution to the deformed RH problem. Related stability, nonzero-boundary and $\bar\partial$-steepest-descent techniques for integrable equations may be found in \cite{Biondini2014,Cuccagna2016,McLaughlin2008}.

We emphasize that the present results extend the finite-genus asymptotic theory for the defocusing NLS equation to a higher-order member of the AKNS hierarchy. Formally, when $\beta=0$, the third-order contribution disappears and the phase function reduces to that of the defocusing NLS equation, whose finite-genus asymptotic theory was obtained in \cite{Fan2026}. When $\beta>0$, the additional Abelian integral $h(z)$ associated with the third-order flow modifies the phase geometry, the critical values, and the local matching data near the transition rays. The analysis developed here may also be useful for studying higher-order AKNS equations of Hirota type, where the phase function is modified by the corresponding higher-order normalized Abelian integrals.

In the following, we state the main asymptotic results and define the four space-time regions determined by the critical values of the phase function.

\subsection{Main results}\label{mainresults}

Let $E_j$ and $\widehat{E}_j,\, (j=0,1,\ldots,n)$ be fixed real-valued constants. We now define $z_j^f,\, z_j^g,\, z_j^h$ in terms of $E_j$ and $\widehat{E}_j$ as follows. Let
\begin{equation} \label{fgh}
	\begin{aligned}
	f(z):=\int_{\widehat{E}_0}^z\frac{\prod_{j=0}^n(s-z_j^f)}{w(s)}\mathrm{d}s,\quad g(z):=\int_{\widehat{E}_0}^z\frac{4\prod_{j=0}^{n+1}(s-z_j^g)}{w(s)}\mathrm{d}s,\quad h(z):=\int_{\widehat{E}_0}^z\frac{12\prod_{j=0}^{n+2}(s-z_j^h)}{w(s)}\mathrm{d}s,
	\end{aligned}
\end{equation}
be three hyperelliptic integrals. The function $w(z)$ determined by (\ref{RSurface}) can make sure that the constants $z_j^f,\, z_j^g,\, z_j^h$ are uniquely determined by the conditions
\begin{equation}
	\begin{cases}
    \begin{array}{llll}
		f(z)=z+\mathcal{O}(1), & \quad  g(z)=2z^2+\mathcal{O}(1), & \quad h(z)=4z^3+\mathcal{O}(1), & z\to\infty, \v\\
		 \int_{E_j}^{\widehat{E}_j}\mathrm{d}f=0, &  \quad \int_{E_j}^{\widehat{E}_j}\mathrm{d}g=0, & \quad  \int_{E_j}^{\widehat{E}_j}\mathrm{d}h=0, & j=1,\ldots,n.
	\end{array}
\end{cases}
\end{equation}
Moreover, $\left(\{z_j^f\}_{j=0}^n\cup\{z_j^g\}_{j=0}^{n+1}\cup\{z_j^h\}_{j=0}^{n+2}\right)\cap\{E_j,\widehat{E}_j\}_{j=0}^n=\emptyset.$

The points $\xi_j,\, \widehat{\xi}_j, (j=1,...,n)$ are precisely the critical points of the phase function $\theta$ defined in \eqref{phasefunction}, and defined by
\begin{equation} \label{xi1}
	\xi_j=-\frac{4\alpha\prod_{k=0}^{n+1}(E_j-z_k^g)+12\beta\prod_{k=0}^{n+2}(E_j-z_k^h)}{\prod_{k=0}^n(E_j-z_k^f)},
\end{equation}
\begin{equation}  \label{xi2}
	\widehat{\xi}_j=-\frac{4\alpha\prod_{k=0}^{n+1}(\widehat{E}_j-z_k^g)+12\beta\prod_{k=0}^{n+2}(\widehat{E}_j-z_k^h)}{\prod_{k=0}^n(\widehat{E}_j-z_k^f)}.
\end{equation}
these points are used to divide the $(x,t)$-half-plane into several distinct space--time regions.

\begin{definition} \label{def1} For any positive constant $C$, set $\xi = x/t$
\begin{itemize}
	\item Transition region I: $\cup_{j=0}^{n}\{\xi : |\xi - \widehat{\xi}_j|t^{2/3} \leq C\}$,
	
	\item Transition region II: $\cup_{j=0}^{n}\{\xi : |\xi - \xi_j|t^{2/3} \leq C\}$,
	
	\item Zakharov--Manakov region III: $\xi \in (-\infty, \widehat{\xi}_n) \cup_{j=1}^{n} (\xi_j, \widehat{\xi}_{j-1}) \cup (\xi_0, +\infty)$,
	
	\item Fast decaying region IV: $\xi \in \cup_{j=0}^{n}(\widehat{\xi}_j, \xi_j)$.
\end{itemize}
\label{def-1}
\end{definition}

\begin{figure}[!t]
	\centering
	\includegraphics[scale=0.5]{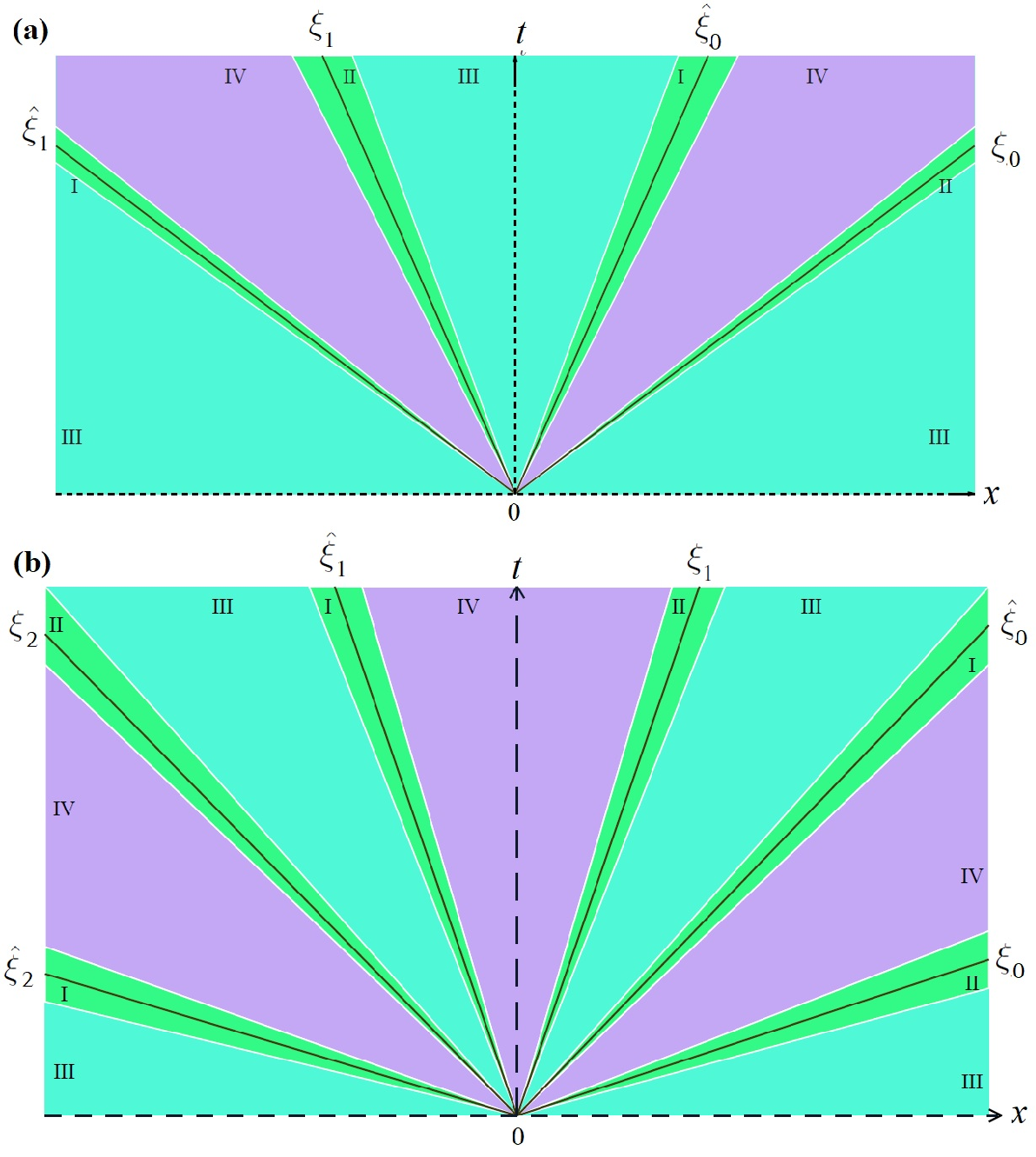}
	\caption{Four distinct asymptotic regions for the different genuses: (a) $n=1$ and (b) $n=2$.}
	\label{fig:Hirota11}
\end{figure}

\begin{remark} \label{rm2} For the general $m$th-order flow equation of the AKNS hierarchy \cite{Ablowitz1974}
 \bee
 iq_t+\sum_{\ell=2}^m\mu_{\ell}\alpha_{\ell}\mathcal{F}_{\ell}(q, q_x, q_{xx},...)=0,
 \ene
 where $\alpha_{\ell}\in\mathbb{R},\, \mu_j=1\, (j=2,4,...), \mu_j=i, (j=3,5,...)$, the NLS double operator $\mathcal{F}_2$, Hirota cubic operator $\mathcal{F}_3$ \cite{Hirota1973}, Lakshmanan-Porsezian-Daniel quartic operator $\mathcal{F}_4$ \cite{4-nls,4-nls2} and quintic operator $\mathcal{F}_5$ \cite{5-nls2, 5-nls} are given by
 \bee \label{h-nls}
 \begin{array}{l}
 \mathcal{F}_2=q_{xx}-2|q|^2q,\v\\
 \mathcal{F}_3=q_{xxx}-6|q|^2q_x,\v\\
  \mathcal{F}_4=q_{xxxx}- 8|q|^2q_{xx}+ 6|q|^4q- 4|q_x|^2q- 6\bar{q}q_x^2-2q^2\bar{q}_{xx}, \v\\
 \mathcal{F}_5=q_{5x}-10|q|^2q_{xxx} -10(q|q_x|^2)_x - 20q^2q_xq_{xx} + 30|q|^4q_x,\v\\
  \cdots \cdots,
 \end{array}
 \ene
  the higher-order hyperelliptic integrals of $g$ and $h$ related to the second- and third-order flows may be chosen as the general form
\bee  \label{hl}
h_{\ell}(z):=\int_{\widehat{E}_0}^z\frac{2^{{\ell}-1}{\ell}\prod_{j=0}^{n+\ell-1}(s-z_j^{h_{\ell}})}{w(s)}\mathrm{d}s,\quad {\ell}=2,3,...,m,
\ene
where $z_j^{h_{\ell}}$'s are uniquely determined by the conditions
\begin{equation}
	\begin{cases}
    \begin{array}{ll}
		 h_{\ell}(z)=2^{{\ell}-1}z^{\ell}+\mathcal{O}(1), & z\to\infty,\,\, {\ell}=2,3,...,m, \v\\
		 \int_{E_j}^{\widehat{E}_j}\mathrm{d}h_{\ell}=0, & j=1,\ldots,n.
	\end{array}
\end{cases}
\end{equation}
Moreover, the phase function (\ref{phasefunction}) in the corresponding RH problem is changed as
\begin{equation}\label{phase-g}
\theta_m(z;\xi)=-(f(z)-f_0)\xi-\sum_{{\ell}=2}^m\alpha_{\ell}(h_{\ell}(z)-h_{{\ell}0}), \quad \xi=\frac{x}{t},
\end{equation}
where $\alpha_{\ell}\, (\ell=2,3,...,m)$ are coefficients of the $m$th-order AKNS flow, and $h_{{\ell}0}$'s defined by
\bee
h_{{\ell}0}:=\lim_{z\to\infty}(h_{\ell}(z)-2^{{\ell}-1}z^{\ell}), \qquad {\ell}=2,3,...,m.
\ene
The points $\xi_j,\, \widehat{\xi}_j, (j=1,...,n)$ are precisely the critical points of the generalized phase function $\theta$ defined in \eqref{phase-g}, and defined by
\begin{equation} \label{xi1g}
	\xi_j=-\frac{\sum_{\ell=2}^{m}\alpha_{\ell}\prod_{k=0}^{n+\ell-1}(E_j-z_k^{h_{\ell}})}{\prod_{k=0}^n(E_j-z_k^f)},
\end{equation}
\begin{equation}  \label{xi2g}
	\widehat{\xi}_j=-\frac{\sum_{\ell=2}^{m}\alpha_\ell\prod_{k=0}^{n+\ell-1}(\widehat{E}_j-z_k^{h_{\ell}})}{\prod_{k=0}^n(\widehat{E}_j-z_k^f)},
\end{equation}
these points are used to divide the $(x,t)$-half-plane into several distinct space--time regions (cf. Definition \ref{def1}).

\end{remark}

An illustration of the four regions for genuses 1 and 2 described above is given in Fig.~\ref{fig:Hirota11}. We derive the long-time asymptotics of $q(x,t)$ in these regions under the following assumptions.

\begin{assumption} Throughout this paper, we impose the following assumptions:
\begin{itemize}
	\item There exists a positive constant $C$ such that $q_0(x)=q^{(AG)}(x,0)$ for $|x|>C$, that is, the initial condition $q_0(x)$ coincides with the finite-genus algebro-geometric background
outside a compact set.
	
	\item Let $r_j\, (j=1,2)$ determined by \eqref{r}, be the reflection coefficients associated with the initial condition $q_0(x)$ and we assume that
$|r_i(z)|=1$ for $z\in\{E_j,\widehat{E}_j\}_{j=0}^n$ and that $r_j$'s are bounded on the set: $\cup_{j=0}^n(E_j,\widehat{E}_j)$.
	
	\item We may take the fixed real numbers $E_j,\, \widehat{E}_j\, (j=0,1,\ldots,n)$, so that the saddle points $z_1$ and $z_2$ defined in \eqref{z} do not simultaneously approach the same critical value $\xi_j$ or $\widehat{\xi}_j$.
\end{itemize}
\label{ass-1}
\end{assumption}

\begin{theorem} \label{theom} For the given finite-genus algebro-geometric solution $q^{(AG)}(x,t)=q^{(AG)}(x,t;\boldsymbol{E},\boldsymbol{\widehat{E}},\boldsymbol{\phi})$ of the defocusing Hirota
equation (\ref{Hirota}), we have the following long-time asympotic behaviors of the Cauchy problem given by (\ref{Hirota}), (\ref{background}) with the constraint (\ref{qRestrict})
and Assumption \ref{ass-1} in the different regions I-IV given by Definition \ref{def-1}:

\begin{itemize}
\item {} {\bf Long-time asymptotics for the transition region I:}
	Let $j_1\in\{0,\ldots,n\}$ be fixed and $|\xi-\widehat{\xi}_{j_1}|t^{2/3}\leq C$, then we have
	\begin{equation}
  q(x,t)=e^{-2\delta(\infty)}q^{(AG)}(x,t;\boldsymbol{E},\boldsymbol{\widehat{E}},\boldsymbol{\phi}-\boldsymbol{\delta})+H_{\widehat{E}_{j_1}}\nu(s)t^{-1/3}
  +\mathcal{O}(t^{-\varrho}),\quad \varrho\in(1/3,2/3),
	\end{equation}
	where the real vector $\boldsymbol\delta=(\delta_1,\ldots,\delta_n)$ and the purely imaginary constant $\delta(\infty)$ are defined in \eqref{delta_1} and \eqref{delta_infty_1}, respectively, $H_{\widehat{E}_{j_1}}$ is defined in \eqref{H_I}, and $\nu(s)$ is given in \eqref{a}.

\item {} {\bf Long-time asymptotics for the transition region II}: Let $j_2\in\{0,\ldots,n\}$ be fixed, and $|\xi-\xi_{j_2}|t^{2/3}\leq C$ then we have
	\begin{equation}
		q(x,t)=e^{-2\delta(\infty)}q^{(AG)}(x,t;\boldsymbol{E},\boldsymbol{\widehat{E}},\boldsymbol{\phi}-\boldsymbol{\delta})+\tilde{H}_{E_{j_2}}\nu(s)t^{-1/3}
+\mathcal{O}(t^{-\varrho}),\quad\varrho\in(1/3,2/3),
	\end{equation}
	where the real vector $\boldsymbol\delta=(\delta_1,\ldots,\delta_n)$ and the purely imaginary constant $\delta(\infty)$ are defined in \eqref{delta_2} and \eqref{delta_infty_2}, respectively, $\tilde{H}_{E_{j_2}}$ is defined in \eqref{H_II}, and $\nu(s)$ is given in \eqref{a}.

\item {} {\bf Long-time asymptotics for the Zakharov-Manakov region III}:	For $\xi\in(-\infty,\widehat{\xi}_n)\cup_{j=1}^n(\xi_j,\widehat{\xi}_{j-1})\cup(\xi_0,+\infty)$,
	\begin{equation}
		\begin{aligned}
			q\left(x,t\right)&=e^{-2\delta(\infty)}q^{(AG)}(x,t;\boldsymbol{E},\boldsymbol{\widehat{E}},\boldsymbol{\phi}-\boldsymbol{\delta}) \\
			& \qquad +\sum_{j=1}^2\frac{ie^{-2\delta(\infty)}}{\sqrt{\theta^{(z_j,2)}(\xi)}}\left(\beta_{21}M_{11}^{(glo)}(z_j)^2-\beta_{12}M_{12}^{(glo)}(z_j)^2\right)t^{-1/2}
+\mathcal{O}(t^{-1}),
		\end{aligned}
	\end{equation}
	where the real vector $\boldsymbol\delta=(\delta_1,\ldots,\delta_n)$ and the purely imaginary constant $\delta(\infty)$ are defined in \eqref{delta_3} and \eqref{delta_infty_3}, respectively, $\theta^{(z_j,2)}(\xi)$ is defined in \eqref{theta_2}, $M^{(glo)}$ is given in \eqref{M_glo}, and $\beta_{12}$ and $\beta_{21}$ are defined in \eqref{beta}.

\item {} {\bf Long-time asymptotics for the fast-decay region IV}: 	For $\xi\in\cup_{j=0}^{n}(\widehat{\xi}_{j},\xi_{j})$,
	\begin{equation}
		\begin{aligned}
			q\left(x,t\right)=e^{-2\delta(\infty)}q^{(AG)}(x,t;\boldsymbol{E},\boldsymbol{\widehat{E}},\boldsymbol{\phi}-\boldsymbol{\delta})+\mathcal{O}(t^{-1}).
		\end{aligned}
	\end{equation}
\end{itemize}

\end{theorem}

\begin{remark} Specifically, when $\alpha\not=0,\,\beta=0$, these results for the defocusing Hirota equation reduces to the corresponding results for the defocusing NLS equation \cite{Fan2026}. When $\alpha=0,\beta\not=0$, these results reduce to the
results for the defocusing cmKdV equation (\ref{mKdVEquation}).  Similar generalizations can also be obtained for higher-order Hirota equations in the AKNS hierarchy by choosing higher orders of $g$ and $h$ (see Remark \ref{rm2}).
\end{remark}

The rest of this paper is organized as follows. In Sec. \ref{Preliminaries}, we recall the finite-genus algebro-geometric solution of the defocusing Hirota equation, construct the associated Jost functions, formulate the RH problems, and analyze the signature table together with the required matrix factorizations. Sections \ref{Region-I} and \ref{Region-II} are devoted to the nonlinear steepest descent analysis in the two transition regions, including the construction of the global parametrix, the local Painlev\'e XXXIV parametrices and the corresponding small-norm RH problems. In Sec. \ref{RegionIII-IV}, we treat the Zakharov-Manakov region-III and the fast-decay region-IV. Finally, we assemble the asymptotic expansions and prove the main theorems in Sec. \ref{Proof-Regions}. Notice that these results can also be extended to  the general $m$th-order flow of the AKNS hierarchy.

\section{Preliminaries}\label{Preliminaries}

\subsection{Algebro-geometric solution via the RH problem}

To analyze the Cauchy problem given by  (\ref{Hirota}) and (\ref{background}) with the constraint (\ref{qRestrict}), we need to construct the algebro-geometric solution $q^{(AG)}$ of the dfocusing Hirota equation (\ref{Hirota}). We first recall some ingredients from the theory of Riemann surfaces~\cite{Belokolos1994,Farkas-Riemann92,Gesztesy2003,Kamvissis2003,Kotlyarov2017}.

Let $\{E_j,\widehat{E}_j\}_{j=0}^n$ be fixed $2(n+1)$ real constants, and $\mathcal{X}$ the Riemann surface of genus $n$ defined by the equation $w^2(z)=P(z)$ with
$w(z)$ given by (\ref{RSurface}),
%\begin{equation}
%	P(z)=\prod_{j=0}^n(z-E_j)(z-\widehat{E}_j),
%\end{equation}
%with
%\begin{equation}
%	E_0<\widehat{E}_0<E_1<\widehat{E}_1<\cdots<E_n<\widehat{E}_n.
%\end{equation}
where $w$ is analytic in $\mathbb{C}\setminus(\cup_{j=0}^n[E_j,\widehat{E}_j])$, with the branch chosen such that $w(z)\sim z^{n+1}$ as $z\to\infty$, and $z=\pi(P)$ denotes the standard projection.
The Riemann surface $\mathcal{X}$ can be constructed as follows. We first glue two sheets along the cuts $(E_j,\widehat{E}_j)$, $j=0,1,\ldots,n$, and then compactify the resulting surface by adding one point at infinity on each sheet.

\begin{figure}[!t]
	\centering
	\includegraphics[scale=0.26]{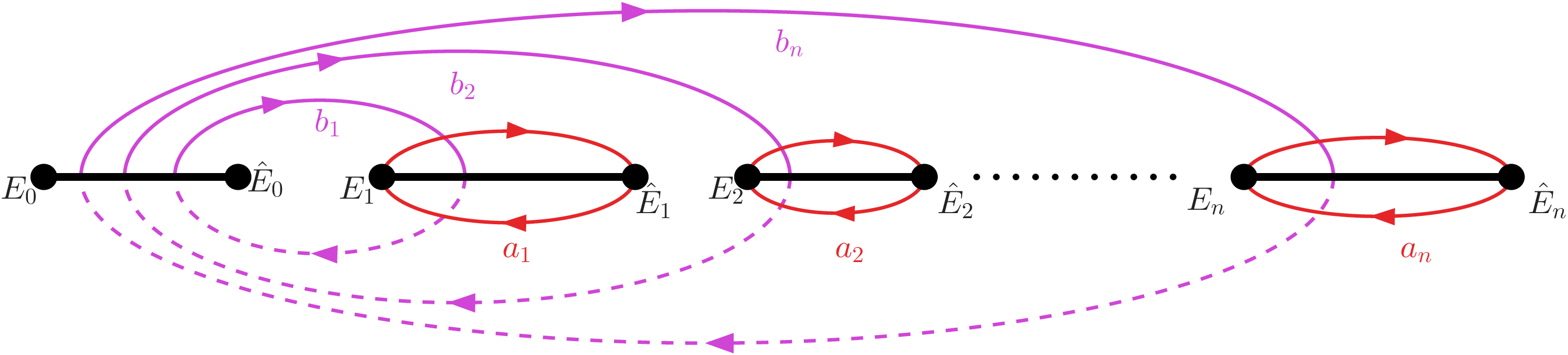}
	\caption{The canonical homology basis $\{a_i,b_i\}_{i=1}^n$ on the Riemann surface $\mathcal{X}$. Here the solid and dashed arcs indicate the parts lying on the upper and lower sheets, respectively.}
	\label{fig:Hirota2}
\end{figure}

The Abelian integrals are defined by
\begin{equation}
	\varpi_j(z)=\int_{\widehat{E}_0}^z\nu_j(s)\,ds,\quad j=1,2,\ldots,n,
\end{equation}
where $d\varpi_j(P)$ form a basis of holomorphic differentials on $\mathcal{X}$:
\begin{equation}
	\nu_j(z)=\frac{\sum_{i=1}^n c_{ji}z^{n-i}}{\sqrt{P(z)}}.
\end{equation}
The coefficients $c_{ji}$ are uniquely determined by the normalization conditions
\begin{equation}
	\oint_{a_i}d\varpi_j(P)=2\int_{E_i}^{\widehat{E}_i}\nu_{j_+}(z)\,dz=\delta_{ji}=\left\{\begin{array}{ll} 1, & i=j,\\ 0, & i\not=j,\end{array}\right.,\quad i,j=1,2,\ldots,n,
\end{equation}
The normalized holomorphic differentials define the $b$-period matrix $B=(B_{ij})_{i,j=1}^n$ by
\begin{equation}
	B_{ij}=\oint_{b_j}d\varpi_i(P)=2\sum_{k=1}^j\int_{\widehat{E}_{k-1}}^{E_k}\nu_i(z)\,dz,
\end{equation}
where the $a_i$-cycle is a closed curve on the upper sheet encircling the interval $(E_i,\widehat{E}_i)$ counterclockwise, $b_j$-cycle starts from $(E_0,\widehat{E}_0)$, runs on the upper sheet to $(E_j,\widehat{E}_j)$, and returns to the starting point on the lower sheet (see Fig.~\ref{fig:Hirota2}). Moreover, $a_i\circ b_j=\delta_{ij},\, (i,j=1,...,n)$.

The matrix $B$ is symmetric and has positive definite imaginary part. Associated with the matrix $B$ is the Riemann theta function defined by the Fourier series
\begin{equation}
	\Theta(\boldsymbol{\jmath})=\sum_{\boldsymbol{m}\in\mathbb{Z}^n}\exp\left(i\pi \boldsymbol{m}^TB\boldsymbol{m}+2i\pi\boldsymbol{m}^T\boldsymbol{\jmath}\right),\quad\boldsymbol{\jmath}=(\jmath_1,\ldots,\jmath_n)\in\mathbb{C}^n,
\end{equation}
together with the column vector of Riemann constants
\begin{equation}
	\begin{aligned}
		\boldsymbol{K}=(K_1,\ldots,K_n),\quad  K_j=\sum_{i=1}^nB_{ji}+\frac{j}{2},\quad  j=1,\ldots,n.
	\end{aligned}
\end{equation}

The theta function is even and satisfies the periodicity relations
\begin{equation}
	\Theta(\boldsymbol{\jmath}\pm \mathbf{e}_j)=\Theta(\boldsymbol{\jmath}),\quad\Theta(\boldsymbol{\jmath}\pm B\mathbf{e}_j)=\Theta(\boldsymbol{\jmath})e^{\mp2\pi i\jmath_j-\pi iB_{jj}},
\end{equation}
where $\mathbf{e}_j=(0,\ldots,0,1,0,\ldots,0)$ is the $j$-th basis vector in $\mathbb{R}^n$. Consequently, the function
\begin{equation}
	\widehat{\Theta}(\boldsymbol{\jmath})=
\frac{\Theta(\boldsymbol{\jmath}+\mathbf{c}+\mathbf{d})}{\Theta(\boldsymbol{\jmath}+\mathbf{d})},\quad \mathbf{c},\mathbf{d}\in\mathbb{R}^n,
\end{equation}
has the periodicity properties
\begin{equation}
	\widehat{\Theta}(\boldsymbol{\jmath}\pm\mathbf{e}_j)=\widehat{\Theta}(\boldsymbol{\jmath}),\quad \widehat{\Theta}(\boldsymbol{\jmath}\pm B\mathbf{e}_j)=e^{\mp2\pi ic_j}\widehat{\Theta}(\boldsymbol{\jmath}).
\end{equation}

The Abel map $\mathcal{A}$ on $\mathcal{X}$, with base point $\hat{P}_0=(0,\widehat{E}_0)$, is defined by
\begin{equation}
	\mathcal{A}(P)=(\mathcal{A}_1(P),\ldots,\mathcal{A}_n(P))^T,\quad \mathcal{A}_j(P)=\int_{\hat{P}_0}^Pd\omega_j(P).
\end{equation}

Based on the functions $f(z)$, $g(z)$, and $h(z)$ defined by (\ref{fgh}), we let
\begin{equation} \label{fgh0}
	f_0:=\lim_{z\to\infty}(f(z)-z),\quad g_0:=\lim_{z\to\infty}(g(z)-2z^2),\quad h_0:=\lim_{z\to\infty}(h(z)-4z^3),
\end{equation}
and introduce the real constants
\begin{equation}
	B_j^f:=\oint_{b_j}df,\quad B_j^g:=\oint_{b_j}dg,\quad B_j^h:=\oint_{b_j}dh,\quad j=1,\ldots,n.
\end{equation}

Let $\boldsymbol{\phi}=(\phi_1,\ldots,\phi_n)\in\mathbb{R}^n$ be  arbitrary, and $\mathcal{D}=\sum_{j=1}^nP_j$ be the unique divisor of degree $n$ such that $P_j$, $j=1,\ldots,n$, lie on the upper sheet of $\mathcal{X}$, with $\pi(P_j)\in(\widehat{E}_{j-1},E_j)$, and satisfy
\begin{equation}
	\prod_{i=0}^n(\pi(P_j)-E_i) =\prod_{i=0}^n(\pi(P_j)-\widehat{E}_i).
\end{equation}

By solving the Riemann-Hilbert problem related to the Baker-Akhiezer function on the Riemann surface \cite{Belokolos1994,Gesztesy2003,Kamvissis2003,Kotlyarov2017,Cao2024}, the finite-genus algebro-geometric solution $q^{(AG)}$ of the defocusing Hirota equation \eqref{Hirota} can be found as
\begin{equation}
	\begin{aligned}
		q^{(AG)}(x,t;\boldsymbol{E},\boldsymbol{\widehat{E}},\boldsymbol{\phi}) & =\frac{i}{2}\left(\sum_{j=1}^n(\widehat{E}_j-E_j)\right)e^{2i(f_0x+(\alpha g_0+\beta h_0)t)} \\
		& \qquad \times\frac{\Theta(\mathcal{A}(\infty)+\mathcal{A}(\mathcal{D})+\boldsymbol{K})\Theta(-\mathcal{A}(\infty)+\boldsymbol{C}(x,t;\boldsymbol{\phi})+\mathcal{A}(\mathcal{D})+\boldsymbol{K})}{\Theta(-\mathcal{A}(\infty)+\mathcal{A}(\mathcal{D})+\boldsymbol{K})\Theta(\mathcal{A}(\infty)+\boldsymbol{C}(x,t;\boldsymbol{\phi})+\mathcal{A}(\mathcal{D})+\boldsymbol{K})},
	\end{aligned}
\label{AG}
\end{equation}
where $\boldsymbol{C}(x,t;\boldsymbol{\phi})=(c_1(x,t;\phi_1),\ldots,c_n(x,t;\phi_n))^T$ is the column vector with components
\begin{equation}
	c_j(x,t;\phi_j)=-\frac{1}{2\pi}\left(B_j^fx+(\alpha B_j^g+\beta B_j^h)t+\phi_j\right),\quad j=1,\ldots,n.
\end{equation}

\begin{remark}  Similarly, we can also generate the finite-genus algebro-geometric solution of the general defocusing $m$th-order equation (\ref{h-nls}) in the AKNS hierarchy
\begin{equation}
	\begin{aligned}
		q_m^{(AG)}(x,t;\boldsymbol{E},\boldsymbol{\widehat{E}},\boldsymbol{\phi}) & =\frac{i}{2}\left(\sum_{j=1}^n(\widehat{E}_j-E_j)\right)e^{2i(f_0x+(\sum_{\ell=2}^m\alpha_{\ell} h_{\ell 0})t)} \\
		& \qquad \times\frac{\Theta(\mathcal{A}(\infty)+\mathcal{A}(\mathcal{D})+\boldsymbol{K})\Theta(-\mathcal{A}(\infty)+\boldsymbol{C}(x,t;\boldsymbol{\phi})+\mathcal{A}(\mathcal{D})+\boldsymbol{K})}{\Theta(-\mathcal{A}(\infty)+\mathcal{A}(\mathcal{D})+\boldsymbol{K})\Theta(\mathcal{A}(\infty)+\boldsymbol{C}(x,t;\boldsymbol{\phi})+\mathcal{A}(\mathcal{D})+\boldsymbol{K})},
	\end{aligned}
\label{AGg}
\end{equation}
where $\boldsymbol{C}(x,t;\boldsymbol{\phi})=(c_1(x,t;\phi_1),\ldots,c_n(x,t;\phi_n))^T$ is the column vector with components
\begin{equation}
	c_j(x,t;\phi_j)=-\frac{1}{2\pi}\left(B_j^fx+(\sum_{\ell=2}^m\alpha_{\ell}B_j^{h_{\ell}})t+\phi_j\right),\quad j=1,\ldots,n.
\end{equation}
with
\begin{equation}
	 B_j^{h_{\ell}}:=\oint_{b_j}dh_{\ell},\quad j=1,\ldots,n.
\end{equation}
\end{remark}

\subsection{Jost functions under the finite-genus algebro-geometric background}

The defocusing Hirota equation \eqref{Hirota} admits the following Lax pair \cite{Ablowitz1974,zhang20}:
\begin{equation}
	\label{lax}
\left\{	\begin{array}{l}
 \Psi_x=X(x,t;z)\Psi,\quad X(x,t;z)=-iz\sigma_3+Q, \v\\
 \Psi_t=T(x,t;z)\Psi,\quad T(x,t;z)=\alpha T_{nls}+\beta T_{cmkdv},
 \end{array}\right.
\end{equation}
where
\begin{equation}
\begin{array}{l}
 Q=Q(x,t)=
	\begin{pmatrix}
		0 & q(x,t) \\
		\overline{q}(x,t) & 0
	\end{pmatrix},\quad
 \sigma_3=\begin{pmatrix}
	1 & 0 \\
	0 & -1
\end{pmatrix}, \v\\
	T_{nls}=2zX+i\sigma_3(Q_x-Q^2),\v\\ T_{cmkdv}=2zT_{nls} +[Q_x,Q]+2Q^3-Q_{xx}.
\end{array}\end{equation}
where $\Psi=\Psi(x,t;z)$ is a matrix-valued eigenfunction, $z\in\mathbb{C}$ a spectral parameter, and  the potential $q$ solves the Hirota equation
(\ref{Hirota}).

For the given finite-genus algebro-geometric  solution $q=q^{(AG)}$ defined by (\ref{AG}), the Jost function $\Psi^{(AG)}$ for the Lax pair \eqref{lax} can be written as \cite{Cao2024,Kotlyarov2017}
\begin{equation}
	\Psi^{(AG)}(x,t;z)=e^{i(f_0x+(\alpha g_0+\beta h_0)t)\sigma_3}\mu^{(AG)}(x,t;z)e^{-i(f(z)x+(\alpha g(x)+\beta h(x))t)\sigma_3},
\end{equation}
where
\begin{equation}
	\mu^{(AG)}(x,t;z)=\frac{1}{2}
	\begin{pmatrix}
		(\kappa(z)+\kappa(z)^{-1})\dfrac{W_{11}(x,t;z,\boldsymbol{\phi})}{W_{11}(\infty;x,t,\boldsymbol{\phi})}&(\kappa(z)-\kappa(z)^{-1})
\dfrac{W_{12}(x,t;z,\boldsymbol{\phi})}{W_{11}(x,t;\infty,\boldsymbol{\phi})} \v\\
		(\kappa(z)-\kappa(z)^{-1})\dfrac{W_{21}(x,t;z,\boldsymbol{\phi})}
{W_{22}(x,t;\infty,\boldsymbol{\phi})}&(\kappa(z)+\kappa(z)^{-1})
\dfrac{W_{22}(x,t;z,\boldsymbol{\phi})}{W_{22}(x,t;\infty,\boldsymbol{\phi})}
	\end{pmatrix}.
\end{equation}
where
\begin{equation}
	\begin{array}{l}
W_{11}(x,t;z,\boldsymbol{\phi})=\dfrac{\Theta(\mathcal{A}(z)+\boldsymbol{C}(x,t;\boldsymbol{\phi})+\mathcal{A}(\mathcal{D})+\boldsymbol{K})}{\Theta(\mathcal{A}(z)+\mathcal{A}(\mathcal{D})+\boldsymbol{K})}, \v\\
W_{12}(x,t;z,\boldsymbol{\phi})=\dfrac{\Theta(-\mathcal{A}(z)+\boldsymbol{C}(x,t;\boldsymbol{\phi})+\mathcal{A}(\mathcal{D})+\boldsymbol{K})}{\Theta(-\mathcal{A}(z)+\mathcal{A}(\mathcal{D})+\boldsymbol{K})}, \v\\
W_{21}(x,t;z,\boldsymbol{\phi})=\dfrac{\Theta(\mathcal{A}(z)+\boldsymbol{C}(x,t;\boldsymbol{\phi})-\mathcal{A}(\mathcal{D})-\boldsymbol{K})}{\Theta(\mathcal{A}(z)-\mathcal{A}(\mathcal{D})-\boldsymbol{K})}, \v\\
W_{22}(x,t;z,\boldsymbol{\phi})=\dfrac{\Theta(\mathcal{A}(z)-\boldsymbol{C}(x,t;\boldsymbol{\phi})+\mathcal{A}(\mathcal{D})+\boldsymbol{K})}{\Theta(\mathcal{A}(z)+\mathcal{A}(\mathcal{D})+\boldsymbol{K})}.
	\end{array}
\end{equation}
and
\begin{equation}
	\kappa(z)=\left(\prod_{j=0}^n\frac{z-E_j}{z-\widehat{E}_j}\right)^{1/4}
\end{equation}
is analytic in $\mathbb{C}\setminus(\cup_{j=0}^n[E_j,\widehat{E}_j])$, with the branch chosen such that $\kappa(\infty)=1$, and
\bee
\kappa_-(z)=i\kappa_+(z),\qquad z\in \cup_{j=0}^n[E_j,\widehat{E}_j], \\
\kappa(z)=\left\{\begin{array}{ll}
  (z-\widehat{E}_j)^{-1/4}+O(1), & z\to \widehat{E}_j, \v\\
  1+\sum_{j=0}^n\frac{\widehat{E}_j-E_j}{4z}+\mathcal{O}(z^{-2}), & z\to \infty
  \end{array}\right.
\ene

\begin{RH} The function $\mu^{(AG)}(x,t;z)$ satisfies the following RH problem \cite{Cao2024,Kotlyarov2017}:
\begin{enumerate}[label=(\textbf{$\mu^{(AG)}$\arabic*}), leftmargin=*]
	\item \textbf{Analyticity:} $\mu^{(AG)}(z)=\mu^{(AG)}(x,t;z)$ is analytic in $\mathbb{C}\setminus(\cup_{j=0}^n[E_j,\widehat{E}_j])$.
	
	\item \textbf{Jump condition:} $\mu^{(AG)}_+(z)=\mu^{(AG)}_-(z)J_{\mu^{(AG)}}(z)$, where the jump matrix is given by ($z\in(E_j,\widehat{E}_j),\,\, j=0,\ldots,n.$)
	\begin{equation} \label{eq:mu^{(AG)}_jump}
		J_{\mu^{(AG)}}(z)=
		\begin{pmatrix}
			0 & -ie^{-i\left(B_j^fx+(\alpha B_j^g+\beta B_j^h)t+\phi_j\right)} \v\\
			-ie^{i\left(B_j^fx+(\alpha B_j^g+\beta B_j^h)t+\phi_j\right)} & 0
		\end{pmatrix}.
	\end{equation}
	
	\item \textbf{Normalization at infinity:} As $z\to\infty$, $\mu^{(AG)}(z)=I+\mathcal{O}(z^{-1})$.
	
	\item \textbf{Local behavior:} $\mu^{(AG)}(z)$ has at most fourth-root singularities near the branch points $E_j$ and $\widehat{E}_j$, $j=0,1,\ldots,n$; that is,
	\begin{equation}
		\mu^{(AG)}(z)=\mathcal{O}((z-p)^{-1/4}),\quad z\to p\in\{E_j,\widehat{E}_j\}_{j=0}^n.
	\end{equation}
\end{enumerate}
\end{RH}
Thus, the algebro-geometric solution $q^{(AG)}$ of (\ref{Hirota}) can be written as
\begin{equation}
	q^{(AG)}(x,t)=2ie^{2i(f_0x+(\alpha g_0+\beta h_0)t)}\lim_{z\to\infty}z\mu_{12}^{(AG)}(z).
\end{equation}

We next consider solutions of the Lax pair \eqref{lax} with $q$ being the solution of the Cauchy problem for the defocusing Hirota equation \eqref{Hirota} with the finite-genus algebro-geometric background \eqref{background}. Let $\mu^{\pm}$ be the solutions of the following Volterra integral equations:
\begin{equation} \label{mu}
	\begin{aligned}
		\mu^\pm(x,t;z) & =e^{i(f_0x+(\alpha g_0+\beta h_0)t)\hat{\sigma}_3}\mu^{(AG)}(x,t;z) \\
		&\qquad  +\int_{\pm\infty}^x\Psi^{(AG)}(x,t;z)\Psi^{(AG)}(s,t;z)^{-1}Q(q-q^{(AG)})(s,t)\mu^\pm(s,t;z)e^{-i(f(z)-f_0)(s-x)\sigma_3}\mathrm{d}s,
	\end{aligned}
\end{equation}
where $	e^{\hat{\sigma}_3}A=e^{\sigma_3}Ae^{-\sigma_3}$
%\end{equation}
%and
%\begin{equation}
%	\begin{aligned}
%		\Gamma(z;s,t,x)&=\Psi^{(AG)}(x,t;z)\Psi^{(AG)}(z;s,t)^{-1} \\
%		& =e^{i(f_0x+(\alpha g_0+\beta h_0)t)\sigma_3}\mu^{(AG)}(x,t;z)e^{if(z)(s-x)\sigma_3}\mu^{(AG)}(z;s,t)^{-1}e^{-i(f_0x+(\alpha g_0+\beta h_0)t)\sigma_3}.
%	\end{aligned}
%\end{equation}

Let $\mu^\pm(x,t;z)=\left(\mu_1^\pm(x,t;z),\mu_2^\pm(x,t;z)\right)$. Then one can easily show that the column vectors $\mu_1^-(x,t;z)$ and $\mu_2^+(x,t;z)$ exist uniquely for $z\in\overline{\mathbb{C}^+}\setminus\{E_j,\widehat{E}_j\}_{j=0}^n$, are analytic in $\mathbb{C}^+$, and admit continuous extensions to $z\in\overline{\mathbb{C}^+}\setminus\{E_j,\widehat{E}_j\}_{j=0}^n$. Similarly, the column vectors $\mu_1^+(x,t;z)$ and $\mu_2^-(x,t;z)$ exist uniquely for $z\in\overline{\mathbb{C}^-}\setminus\{E_j,\widehat{E}_j\}_{j=0}^n$, are analytic in $\mathbb{C}^-$, and admit continuous extensions to $z\in\overline{\mathbb{C}^-}\setminus\{E_j,\widehat{E}_j\}_{j=0}^n$.  Moreover, $(\mu_1^{\mp}(z), \mu_2^{\pm}(z))=I+\mathcal{O}(z^{-1})$ as $z\to \infty$.

Next, we define
\begin{equation} \label{psi}
	\Psi^\pm(x,t;z):=\mu^\pm(x,t;z)e^{-i[(f(z)-f_0)x+(\alpha(g(z)-g_0) +\beta(h(z)-h_0))t]\sigma_3},
\end{equation}
which are two linearly independent solutions of the Lax pair \eqref{lax}. Thus, there exists a scattering matrix $S(z)=(S_{ij}(z))_{2\times 2}$, independent of $(x,t)$, such that
\begin{equation} \label{S}
	\Psi^+(x,t;z)=\Psi^-(x,t;z)S(z), \quad z\in\mathbb{R}\setminus(\cup_{j=0}^n[E_j,\widehat{E}_j]),
\end{equation}
and $\det S(z)=1$ for $z\in\mathbb{R}\setminus(\cup_{j=0}^n[E_j,\widehat{E}_j])$. It follows from (\ref{psi}) and (\ref{S}) that
\bee \label{rel}
\begin{array}{l}
 S_{11}(z)=|\Psi_1^+(x,t;z), \Psi_2^-(x,t;z)|=|\mu_1^+(x,t;z), \mu_2^-(x,t;z)|, \v\\
 S_{22}(z)=|\Psi_1^-(x,t;z), \Psi_2^+(x,t;z)|=|\mu_1^-(x,t;z), \mu_2^+(x,t;z)|, \v\\
  S_{12}(z)=|\Psi_2^+(x,t;z), \Psi_2^-(x,t;z)|=|\mu_2^+(x,t;z), \mu_2^-(x,t;z)|e^{2i[(f(z)-f_0)x+(\alpha(g(z)-g_0) +\beta(h(z)-h_0))t]}, \v\\
   S_{21}(z)=|\Psi_1^-(x,t;z), \Psi_1^+(x,t;z)|=|\mu_1^-(x,t;z), \mu_1^+(x,t;z)|e^{-2i[(f(z)-f_0)x+(\alpha(g(z)-g_0) +\beta(h(z)-h_0))t]}.
\end{array} \ene
Similarly, based on (\ref{rel}), the scattering data $S_{11}(z)$ and $S_{22}(z)$ are analytic in $\mathbb{C}^-$ and $\mathbb{C}^+$, respectively, and admit continuous extensions to $\overline{\mathbb{C}^-}\setminus\{E_j,\widehat{E}_j\}_{j=0}^n$ and $\overline{\mathbb{C}^+}\setminus\{E_j,\widehat{E}_j\}_{j=0}^n$, respectively.
In fact, $S_{i(3-i)}\, (i=1,2)$ admit continuous extensions to $\overline{\mathbb{R}}\setminus\{E_j,\widehat{E}_j\}_{j=0}^n$.
For $z\in \cup_{j=0}^n(E_j,\widehat{E}_j)$,
\bee \label{rel2}
\begin{array}{l}
   S_{12}(z)=|\mu_2^+e^{i[(f_+(z)-f_0)x+(\alpha(g_+(z)-g_0) +\beta(h_+(z)-h_0))t]}, \mu_2^-e^{i[(f_-(z)-f_0)x+(\alpha(g_-(z)-g_0) +\beta(h_-(z)-h_0))t]}|, \v\\
   S_{21}(z)=|\mu_1^-e^{-i[(f_+(z)-f_0)x+(\alpha(g_+(z)-g_0) +\beta(h_+(z)-h_0))t]}, \mu_1^+e^{-i[(f_-(z)-f_0)x+(\alpha(g_-(z)-g_0) +\beta(h_-(z)-h_0))t]}|,
\end{array} \ene

 The remaining entries generally do not admit analytic continuations to $\mathbb{C}^{\pm}$. Moreover, $S_{11}(z),\, S_{22}(z)=1+\mathcal{O}(z^{-1}),\, S_{12}(z),\, S_{21}(z)=\mathcal{O}(z^{-1})$ as $z\to \infty$.

Moreover, we have the following symmetry relations:
\begin{equation}
	S_{11}(z)=\overline{S_{22}(z)},\quad S_{12}(z)=\overline{S_{21}(z)},\quad z\in\mathbb{R}\setminus(\cup_{j=0}^n[E_j,\widehat{E}_j]).
\end{equation}
and $|S_{11}(z)|^2-|S_{12}(z)|^2=1$. We define the reflection coefficients in terms of the scattering matrix $S(z)$ by
\begin{equation}
	\label{r}
	r_1(z)=\frac{S_{12}(z)}{S_{11}(z)},\quad r_2(z)=\frac{S_{21}(z)}{S_{11}(z)}, \quad z\in\mathbb{R}\setminus(\cup_{j=0}^n[E_j,\widehat{E}_j]),
\end{equation}
which generates
\begin{equation}
	\label{r}
	\overline{r_1(z)}=\frac{S_{21}(z)}{S_{22}(z)},\quad \overline{r_2(z)}=\frac{S_{12}(z)}{S_{22}(z)}, \quad z\in\mathbb{R}\setminus(\cup_{j=0}^n[E_j,\widehat{E}_j]).
\end{equation}
and $ |r_1(z)|^2=|r_2(z)|^2=1-|s_{11}|^{-2}.$

\subsection{Two types of Riemann-Hilbert problems and properties}

\subsubsection{Two types of Riemann-Hilbert problems}

According the analyticity of Jost functions $\mu_j^\pm(x,t;z)\, (j=1,2)$ defined by (\ref{mu}) and scattering data $S_{jj}\, (j=1,2)$ given by (\ref{S}), we introduce these two types of matrix-valued functions
\begin{equation}
	\begin{aligned}
		M(x,t;z) & =
		\begin{cases}
			\left(\dfrac{\mu_1^-(x,t;z)}{S_{22}(z)},\mu_2^+(x,t;z)\right),\quad & z\in\mathbb{C}^+, \\[1.2em]
			\left(\mu_1^+(x,t;z),\dfrac{\mu_2^-(x,t;z)}{S_{11}(z)}\right),\quad & z\in\mathbb{C}^-,
		\end{cases} \\
		N(x,t;z) & =
		\begin{cases}
			\left(\mu_1^-(x,t;z),\dfrac{\mu_2^+(x,t;z)}{S_{22}(z)}\right),\quad & z\in\mathbb{C}^+, \\[1.2em]
			\left(\dfrac{\mu_1^+(x,t;z)}{S_{11}(z)},\mu_2^-(x,t;z)\right),\quad & z\in\mathbb{C}^-.
		\end{cases}
	\end{aligned}
\end{equation}

\begin{RH} \label{M}
$M(x,t;z)$ satisfies the following RH problem:

\begin{enumerate}[label=(\textbf{$M$\arabic*}), leftmargin=*]
	\item \textbf{Analyticity:} $M(z)$ is analytic in $\mathbb{C}\setminus\mathbb{R}$.
	
	\item \textbf{Jump condition:} $M_+(z)=M_-(z)J_{M}(z)$ for $z\in\mathbb{R}$, where the jump matrix is given by
	\begin{equation} \label{eq:M_jump}
		J_{M}(z)=
		\begin{cases}
			-ie^{i(f_0x+(\alpha g_0+\beta h_0)t-(B_j^fx+(\alpha B_j^g+\beta B_j^h)t+\phi_j)/2)\hat{\sigma}_3}\sigma_1,
			 %\begin{pmatrix}				0 & 1 \\				1 & 0			\end{pmatrix}
&z\in(E_j,\widehat{E}_j),j=0,\ldots,n, \\
			\\
			\begin{pmatrix}
				1-|r_1(z)|^2 & r_1(z)e^{2it\theta(z)} \v\\
				-\overline{r_1(z)}e^{-2it\theta(z)} & 1
			\end{pmatrix},& z\in\mathbb{R}\setminus(\cup_{j=0}^n[E_j,\widehat{E}_j]),
		\end{cases}
	\end{equation}
	where $\sigma_1=\begin{pmatrix}				0 & 1 \\				1 & 0			\end{pmatrix}$ is the  Pauli matrix, $\theta(z)=\theta(z;\xi)$ is given by (\ref{phasefunction}), i.e.,
	\begin{equation}
		%\label{theta}
		\theta(z)=\theta(z;\xi)=-(f(z)-f_0)\xi-\alpha(g(z)-g_0) -\beta(h(z)-h_0),\quad\xi=\frac{x}{t}.
	\end{equation}
	
	\item \textbf{Normalization at infinity:} As $z\to\infty$, $M(z)=I+\mathcal{O}(z^{-1})$.
	
	\item \textbf{Local behavior:} For $p\in\{E_j,\widehat{E}_j\}_{j=0}^n$, $M(z)$ has the following local behavior:
	\begin{equation}
		\begin{aligned}
			M(z)=(\mathcal{O}((z-p)^{\pm 1/4}),\mathcal{O}((z-p)^{\mp 1/4})),\quad z\to p\mathrm{~from~}\mathbb{C}^{\pm}.
			%M(z)=(\mathcal{O}((z-p)^{-1/4}),\mathcal{O}((z-p)^{1/4})),\quad z\to p\mathrm{~from~}\mathbb{C}^-.
		\end{aligned}
	\end{equation}
\end{enumerate}
\end{RH}

\begin{RH} \label{N} $N(x, t;z)$ satisfies the following Riemann--Hilbert problem:

\begin{enumerate}[label=(\textbf{$N$\arabic*}), leftmargin=*]
	\item \textbf{Analyticity:} $N(z)$ is analytic in $\mathbb{C}\setminus\mathbb{R}$.
	
	\item \textbf{Jump condition:}$N_+(z)=N_-(z)J_{N}(z)$ for $z\in\mathbb{R}$, where the jump matrix is given by
	\begin{equation} \label{eq:N_jump}
		J_{N}(z)=
		\begin{cases}
			-ie^{i(f_0x+(\alpha g_0+\beta h_0)t-(B_j^fx+(\alpha B_j^g+\beta B_j^h)t+\phi_j)/2)\hat{\sigma}_3}\sigma_1,
			&z\in(E_j,\widehat{E}_j),j=0,\ldots,n, \\
			\\
			\begin{pmatrix}
				1 & \overline{r_2(z)}e^{2it\theta(z)} \v\\
				-r_2(z)e^{-2it\theta(z)} & 1-|r_2(z)|^2
			\end{pmatrix},& z\in\mathbb{R}\setminus(\cup_{j=0}^n[E_j,\widehat{E}_j]),
		\end{cases}
	\end{equation}
	
	\item \textbf{Normalization at infinity:} As $z\to\infty$, $N(z)=I+\mathcal{O}(z^{-1})$.
	
	\item \textbf{Local behavior:} For $p\in\{E_j,\widehat{E}_j\}_{j=0}^n$, $N(z)$ has the following local behavior:
	\begin{equation}
		\begin{aligned}
			N(z)=(\mathcal{O}((z-p)^{\mp 1/4}),\mathcal{O}((z-p)^{\pm 1/4})),\quad z\to p\mathrm{~from~}\mathbb{C}^{\pm}.
		%	N(z)=(\mathcal{O}((z-p)^{1/4}),\mathcal{O}((z-p)^{-1/4})),\quad z\to p\mathrm{~from~}\mathbb{C}^-.
		\end{aligned}
	\end{equation}
\end{enumerate}
\end{RH}

We have the following reconstruction formula between the solution $q(x,t)$ and $M(z)$ or $N(z)$:
\begin{equation}\label{inversion formula}
	q(x,t)=2i\lim_{z\to\infty}zM_{12}(z)=2i\lim_{z\to\infty}zN_{12}(z).
\end{equation}

\subsubsection{Factorizations of the jump matrices}

For the later use, we record several factorizations of the jump matrices $J_M$ and $J_N$, which will be used to deform the contours onto the steepest descent paths determined by the signature table of the phase function $\theta$.

We have the following factorizations:
\begin{equation}
	\begin{aligned}
		J_M(z) & =
		\begin{pmatrix}
			1 & 0 \v \\
			-\frac{\overline{r_1(z)}e^{-2it\theta(z)}}{1-|r_1(z)|^2} & 1
		\end{pmatrix}(1-|r_1(z)|^2)^{\sigma_3}
		\begin{pmatrix}
			1 & \frac{r_1(z)e^{2it\theta(z)}}{1-|r_1(z)|^2} \v\\
			0 & 1
		\end{pmatrix} \\
		& =
		\begin{pmatrix}
			1 & r_1(z)e^{2it\theta(z)} \v\\
			0 & 1
		\end{pmatrix}
		\begin{pmatrix}
			1 & 0 \v\\
			-\overline{r_1(z)}e^{-2it\theta(z)} & 1
		\end{pmatrix},\quad z\in\mathbb{R}\setminus(\cup_{j=0}^n[E_j,\widehat{E}_j]),
	\end{aligned}
\end{equation}
and
\begin{equation}
\begin{aligned}
		J_M\left(z\right) &=
		\begin{pmatrix}
			1 & r_1(z)e^{2it\theta_-(z)} \v\\
			0 & 1
		\end{pmatrix}
		\begin{pmatrix}
			0 & -ie^{i(t(\theta_+(z)+\theta_-(z))-\phi_j)} \v\\
			-ie^{-i(t(\theta_+(z)+\theta_-(z))-\phi_j)} & 0
		\end{pmatrix} \\
		&\quad\quad\quad \times
		\begin{pmatrix}
			1 & 0 \v\\
			-r_1(z)e^{-2i(t\theta_+(z)-\phi_j)} & 1
		\end{pmatrix} \\
		&  =
		\begin{pmatrix}
			1 & 0 \v\\
			-\frac{r_1(z)e^{-2i(t\theta_-(z)-\phi_j)}}{1-(r_1(z))^2e^{2i\phi_j}} & 1
		\end{pmatrix}
		\begin{pmatrix}
			0 & -ie^{i(t(\theta_+(z)+\theta_-(z))-\phi_j)} \v\\
			-ie^{-i(t(\theta_+(z)+\theta_-(z))-\phi_j)} & 0
		\end{pmatrix} \\
		& \quad\quad\quad \times
		\begin{pmatrix}
			1 & \frac{r_1(z)e^{2it\theta_+(z)}}{1-(r_1(z))^2e^{2i\phi_j}} \v\\
			0 & 1
		\end{pmatrix},\quad z\in\cup_{j=0}^n(E_j,\widehat{E}_j),
	\end{aligned}
\end{equation}
where $\theta_+(z)+\theta_-(z)=2(f_0\xi+ \alpha g_0+\beta h_0)-B_j^f \xi- \alpha B_j^g-\beta B_j^h$.

Similarly, for the jump matrix $J_N$, one has
\begin{equation}
	\begin{aligned}
		J_N(z) & =
		\begin{pmatrix}
			1 & \frac{r_2(z)e^{2it\theta(z)}}{1-|r_2(z)|^2} \v\\
			0 & 1
		\end{pmatrix}(1-|r_2(z)|^2)^{\sigma_3}
		\begin{pmatrix}
			1 & 0 \v\\
			-\frac{\overline{r_2(z)}e^{-2it\theta(z)}}{1-|r_2(z)|^2} & 1
		\end{pmatrix}  \v\\
		& =
		\begin{pmatrix}
			1 & 0 \v\\
			-r_2(z)e^{-2it\theta(z)} & 1
		\end{pmatrix}
		\begin{pmatrix}
			1 & \overline{r_2(z)}e^{2it\theta(z)}  \v\\
			0 & 1
		\end{pmatrix},\quad z\in\mathbb{R}\setminus(\cup_{j=0}^n[E_j,\widehat{E}_j]),
	\end{aligned}
\end{equation}
and
\begin{equation}
	\begin{aligned}
		J_N\left(z\right)& =
		\begin{pmatrix}
			1 & 0 \v\\
			-r_2(z)e^{-2it\theta_-(z)} & 1
		\end{pmatrix}
		\begin{pmatrix}
			0 & -ie^{i(t(\theta_+(z)+\theta_-(z))-\phi_j)} \v\\
			-ie^{-i(t(\theta_+(z)+\theta_-(z))-\phi_j)} & 0
		\end{pmatrix} \v\\
		& \qquad \times \begin{pmatrix}
			1 & r_2(z)e^{2i(t\theta_+(z)-\phi_j)} \v\\
			0 & 1
		\end{pmatrix} \v\\
		& =
		\begin{pmatrix}
			1 & \frac{r_2(z)e^{2i(t\theta_-(z)-\phi_j)}}{1-(r_2(z))^2e^{-2i\phi_j}}  \v\\
			0 & 1
		\end{pmatrix}
		\begin{pmatrix}
			0 & -ie^{i(t(\theta_+(z)+\theta_-(z))-\phi_j)} \v\\
			-ie^{-i(t(\theta_+(z)+\theta_-(z))-\phi_j)} & 0
		\end{pmatrix} \v\\
		&\qquad \qquad \times
		\begin{pmatrix}
			1 & -\frac{r_2(z)e^{-2it\theta_+(z)}}{1-(r_2(z))^2e^{-2i\phi_j}} \v\\
			0 & 1
		\end{pmatrix},\quad z\in\cup_{j=0}^n(E_j,\widehat{E}_j).
	\end{aligned}
\end{equation}

\subsubsection{Signature table for $\operatorname{Im}\theta$}

For the given phase $\theta(z; \xi)$ in (\ref{phasefunction}), a point $z=z(\xi)\in\mathbb{R}$ is called its saddle point if it satisfies one of the following conditions:
\begin{equation}\label{z}
	\begin{aligned}
		& (a)\quad\theta^{\prime}(z; \xi)=0,&\mathrm{~for~}z\in\mathbb{R}\backslash(\cup_{j=0}^n(E_j,\widehat{E}_j)); \\
		& (b)\quad\operatorname{Im}\theta(z; \xi)=0,&\mathrm{~for~}z\in\cup_{j=0}^n(E_j,\widehat{E}_j).
	\end{aligned}
\end{equation}

It follows from \eqref{phasefunction} and (\ref{fgh}) that
\begin{equation}
\theta^{\prime}(z; \xi)=-\frac{1}{w(z)}\left[\xi\prod_{j=0}^n(z-z_j^f)+4\alpha\prod_{j=0}^{n+1}(z-z_j^g)+12\beta\prod_{j=0}^{n+2}(z-z_j^h)\right],
\end{equation}
so that its zeros are precisely the roots of
\begin{equation}
	F(z;\xi)=\xi\prod_{j=0}^n(z-z_j^f)+4\alpha\prod_{j=0}^{n+1}(z-z_j^g)+12\beta\prod_{j=0}^{n+2}(z-z_j^h)=0.
\end{equation}

Since $f(E_j)=f(\widehat{E}_j)$, $g(E_j)=g(\widehat{E}_j)$, and $h(E_j)=h(\widehat{E}_j)$ for $j=0,1,\ldots,n$, we have $\theta(E_j)=\theta(\widehat{E}_j)$. This implies that $F(z;\xi)$ has at least one zero in each interval $(E_j,\widehat{E}_j)$. Since $F(z;\xi)$ is a polynomial of degree $n+3$, there remain two zeros. If such a zero lies outside the union of these intervals, namely in $\mathbb{R}\setminus(\cup_{j=0}^n(E_j,\widehat{E}_j))$, then it corresponds to case $(a)$. On the other hand, if such a zero lies inside one of the intervals $(E_j,\widehat{E}_j)$, then it corresponds to case $(b)$. This gives rise to two saddle points, denoted by $z_1$ and $z_2$ (without loss of generality, assume that $z_1<z_2$).

We claim that the two saddle points cannot coincide. Indeed, since
\begin{equation}
	\theta^{\prime\prime}(z; \xi)=\frac{F(z;\xi)w^{\prime}(z)-\partial_{z}F(z;\xi)w(z)}{w^2(z)},
\end{equation}
When $\theta^{\prime}(z; \xi)=0$, noting that $\alpha,\beta>0$ do not affect the sign, this reduces to
\begin{equation}
	\theta^{\prime\prime}(z; \xi)=-\frac{\partial_{z}F(z;\xi)}{w(z)}<0.
\end{equation}

For case $(a)$, let $z_1=z_1(\xi)\in\mathbb{R}\setminus(\cup_{i=0}^n(E_i,\widehat{E}_i))$. In view of the identities
\begin{equation}
	F(E_j;\xi_j)=F(\widehat{E}_j;\widehat{\xi}_j)=0,\quad j=0,1,\ldots,n,
\end{equation}
where $\xi_j,\, \widehat{\xi}_j$ are given by (\ref{xi1}) and (\ref{xi2}).
%\begin{equation}
%	\xi_j=-\frac{4\alpha\prod_{k=0}^{n+1}(E_j-z_k^g)+12\beta\prod_{k=0}^{n+2}(E_j-z_k^h)}{\prod_{k=0}^n(E_j-z_k^f)},
%\end{equation}
%\begin{equation}
%	\widehat{\xi}_j=-\frac{4\alpha\prod_{k=0}^{n+1}(\widehat{E}_j-z_k^g)+12\beta\prod_{k=0}^{n+2}(\widehat{E}_j-z_k^h)}{\prod_{k=0}^n(\widehat{E}_j-z_k^f)}.
%\end{equation}

By the implicit function theorem, taking the interval $(\widehat{E}_j,E_{j+1})$ as an example, we obtain
\begin{equation}
	\partial_\xi z_1(\xi)=-\frac{\partial_\xi F(z_1;\xi)}{\partial_{z_1}F(z_1;\xi)}=-\frac{\prod_{k=0}^n(z_1-z_k^f)}{\partial_{z_1}F(z_1;\xi)}<0.
\end{equation}
Indeed, if $n-j$ is even, then both $\partial_{z_1}F(z_1;\xi)$ and $\prod_{k=0}^n(z_1-z_k^f)$ are positive; if $n-j$ is odd, then both quantities are negative. Hence $\partial_\xi z_1(\xi)<0$. Therefore, $z_1(\xi)$ is a monotonically decreasing function of $\xi$ on each interval of $\mathbb{R}\setminus(\cup_{i=0}^n(E_i,\widehat{E}_i))$.

Similarly, in case $(b)$, $z_1(\xi)$ is a monotonically decreasing function of $\xi$ on $\cup_{j=0}^n(E_j,\widehat{E}_j)$. Moreover, $z_1(\xi)$ tends to $\widehat{E}_j$ as $\xi$ tends to $\widehat{\xi}_j$. The same conclusion holds for $z_2=z_2(\xi)$; in particular, we have $z_1(\widehat{\xi}_j)=\widehat{E}_j$ and $z_2(\xi_j)=E_j$.

\begin{remark} For the general $m$th-order flow of the AKNS hierarchy, it follows from (\ref{fgh}), (\ref{hl}), (\ref{phase-g}) that \begin{equation}
\theta_m^{\prime}(z; \xi)=-\frac{1}{w(z)}\left[\xi\prod_{j=0}^n(z-z_j^f)
+\sum_{{\ell}=2}^m\alpha_{\ell} 2^{{\ell}-1}{\ell}\prod_{j=0}^{n+\ell-1}(z-z_j^{h_{\ell}})\right],
\end{equation}
so that its zeros are precisely the roots of
\bee
 F_m(z;\xi)=\xi\prod_{j=0}^n(z-z_j^f)
+\sum_{{\ell}=2}^m\alpha_{\ell} 2^{{\ell}-1}{\ell}\prod_{j=0}^{n+\ell-1}(z-z_j^{h_{\ell}}).
\ene
Similarly, one can also give the signature table for $\operatorname{Im}\theta$ in principle.
\end{remark}

\section{Long-time asymptotics in the transition region I}\label{Region-I}

In this section, we carry out the asymptotic analysis of the Riemann--Hilbert problem \ref{M} for $M(x,t;z)$ in order to derive the asymptotics of $q$ in the transition region I. According to Definition~\ref{def1}, we may choose $C$ so that, throughout this section, only one of the saddle points $z_1$ is involved in the transition regime
$-C\le(\xi-\widehat{\xi}_{j_1})t^{2/3}\leq0,$ for some fixed $j_1\in\{0,\ldots,n\}$, since the analysis in the complementary half-region is analogous. For large $t$, one has $\xi_{j_1+1}<\xi<\widehat{\xi}_{j_1}$, and hence the saddle point satisfies $z_1\in[\widehat{E}_{j_1},E_{j_1+1})$.

Let $z_{j_1}\in(E_{j_1},\widehat{E}_{j_1})$ be fixed. We set
\begin{equation}
	\begin{aligned}
		&\Omega_1=\Omega_1(\xi)=\{z\in\mathbb{C}:0\leq\arg(z-z_1)\leq\varphi_1\},\quad \Sigma_1=\Sigma_1(\xi)=z_1+e^{i\varphi_1}\mathbb{R}^+,\\
		&\Omega_2=\Omega_2(\xi)=\{z\in\mathbb{C}:\pi-\varphi_1\leq\arg(z-z_{j_1})\leq\pi\},\quad \Sigma_2=\Sigma_2(\xi)=z_{j_1}+e^{i(\pi-\varphi_1)}\mathbb{R}^+,
	\end{aligned}
\end{equation}
where $\varphi_{1}$ is chosen such that
\begin{equation}
	\operatorname{Im}\theta(z)
	\begin{cases}
		<0,\qquad  z\in\Sigma_1\setminus\{z_1\}, \\
		>0, \qquad z\in\Sigma_2\setminus\{z_{j_1}\},
	\end{cases}
\end{equation}
and $U^*$ denotes the complex conjugate of a region $U$.

The first transformation of $M$ is defined by
\begin{equation}
	M^{(1)}(z)=e^{\delta(\infty)\sigma_3}M(z)G_M(z)e^{-\delta(z)\sigma_3},
\end{equation}
where
\begin{equation}
	G_M(z)=
	\begin{cases}
		\begin{pmatrix}
			1 & 0 \v\\
			\overline{r_1(z)}e^{-2it\theta(z)} & 1
		\end{pmatrix},\quad z\in\Omega_1, \v\\
		\begin{pmatrix}
			1 & r_1(z)e^{2it\theta(z)} \v\\
			0 & 1
		\end{pmatrix},\quad z\in\Omega_1^*, \v\\
		\begin{pmatrix}
			1 & -\frac{r_1(z)e^{2it\theta(z)}}{1-|r_1(z)|^2} \v\\
			0 & 1
		\end{pmatrix},\quad z\in\Omega_2, \v\\
		\begin{pmatrix}
			1 & 0 \v\\
			-\frac{\overline{r_1(z)}e^{-2it\theta(z)}}{1-|r_1(z)|^2} & 1
		\end{pmatrix},\quad z\in\Omega_2^*, \v\\
		I, \quad \text{elsewhere,}
	\end{cases}
\end{equation}
is a matrix-valued function determined by the factorizations of the jump matrix $J_M$. The scalar function $\delta(z)$ is defined by
\begin{equation}
	\delta(z)=\frac{w(z)}{2\pi i}\left[\sum_{j=1}^n\delta_j\int_{E_j}^{\widehat{E}_j}\frac{i\mathrm{~d}s}{w_+(s)(s-z)}-\int_{(-\infty,E_{j_1})\setminus(\cup_{j=0}^{j_1-1}(E_j,\widehat{E}_j))}\frac{\log(1-|r_1(s)|^2)\mathrm{d}s}{w(s)(s-z)}\right],
\end{equation}
where the logarithm is taken on the principal branch, and the constants $\delta_j$, $j=1,\ldots,n$, are determined by the linear system
\begin{equation}\label{delta_1}
	\int_{(-\infty,E_{j_1})\setminus(\cup_{j=0}^{j_1-1}(E_j,\widehat{E}_j))}\frac{\log(1-|r_1(s)|^2)s^k\mathrm{d}s}{w(s)}
=i\sum_{j=1}^n\delta_j\int_{E_j}^{\widehat{E}_j}\frac{s^k\mathrm{d}s}{w_+(s)},\quad k=0,\ldots,n-1.
\end{equation}

Here it is understood that $(E_{j_1-1},\widehat{E}_{j_1-1})=\emptyset$ if $j_1=0$. In addition, one can prove that each $\delta_j$ is real.

\begin{RH} $\delta(z)$ satisfies the following Riemann--Hilbert problem:
\begin{enumerate}[label=(\textbf{$\delta$\arabic*}), leftmargin=*]
	\item \textbf{Analyticity:} $\delta(z)$ is analytic in $\mathbb{C}\setminus(\cup_{j=0}^n[E_j,\widehat{E}_j]\cup(-\infty,\widehat{E}_{j_1}])$.
	
	\item \textbf{Jump condition:} $\delta(z)=-\overline{\delta(\bar{z})}$ satisfies the jump relation
	\begin{equation} \label{eq:delta_jump}
		\left\{\begin{aligned}
			&\delta_-(z)  =\delta_+(z)+\log(1-|r_1(z)|^2),\quad &z\in(-\infty,E_{j_1})\setminus(\cup_{j=0}^{j_1-1}[E_j,\widehat{E}_j]), \\
			 & \delta_-(z)=-\delta_+(z)+i\delta_j,\quad &z\in(E_j,\widehat{E}_j),\quad j=1,\ldots,n.
		\end{aligned}\right.
	\end{equation}
	
	\item \textbf{Normalization at infinity:} As $z \to \infty$,
	$\delta(z)=\delta(\infty)+\frac{\delta^{(1)}}{z}+\mathcal{O}(z^{-2}),$
		where
	\begin{equation}\label{delta_infty_1}
		\begin{aligned}
			\delta(\infty) =\frac{1}{2\pi i}\left[\int_{(-\infty,E_{j_1})\setminus(\cup_{j=0}^{j_1-1}(E_j,\widehat{E}_j))}\frac{\log(1-|r_1(s)|^2)s^n \mathrm{d}s}{w(s)} -i\sum_{j=1}^n\delta_j\int_{E_j}^{\widehat{E}_j}\frac{s^n\mathrm{d}s}{w_+(s)}\right],
		\end{aligned}
	\end{equation}
	and
	\begin{equation}
		\begin{aligned}
			&\delta^{(1)} =\delta(\infty)\sum_{j=0}^n(E_j+\widehat{E}_j) \\
			&\qquad\quad  -\frac{1}{2\pi i}\left[\int_{(-\infty,E_{j_1})\setminus(\cup_{j=1}^{j_1-1}(E_j,\widehat{E}_j))}\frac{\log(1-|r_1(s)|^2)s^{n+1} \mathrm{d}s}{w(s)}-i\sum_{j=1}^n\delta_j\int_{E_j}^{\widehat{E}_j}\frac{s^{n+1}\mathrm{d}s}{w_+(s)}\right].
		\end{aligned}
	\end{equation}
	
	\item \textbf{Local behavior:} As $z\to p\in\{E_j\}_{j=0}^{j_1}\cup\{\widehat{E}_j\}_{j=0}^{j_1-1}$ from $\mathbb{C}^+$, we have
	$e^{\delta(z)}=\mathcal{O}((z-p)^{1/2}),$
	and
	$		\delta(z)=\frac{i}{2}\delta_{j_1}+\mathcal{O}((z-\widehat{E}_{j_1})^{1/2}),\quad z\to\widehat{E}_{j_1} \text{~from~} \mathbb{C}\setminus(-\infty,\widehat{E}_{j_1}).$
\end{enumerate}
\end{RH}

\begin{figure}[!t]
	\centering
	\includegraphics[scale=0.33]{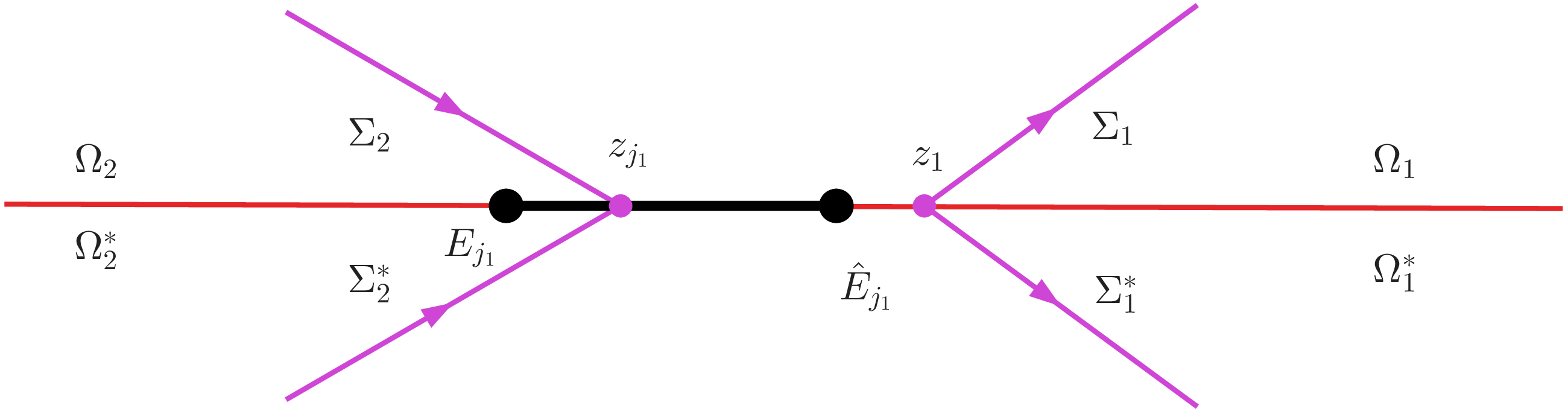}
	\caption{The contours $\Sigma^{(1)}$.}
	\label{fig:Hirota3}
\end{figure}

\begin{RH}  $M^{(1)}$ satisfies the following Riemann-Hilbert problem:

\begin{enumerate}[label=(\textbf{$M^{(1)}$\arabic*}), leftmargin=*]
	\item \textbf{Analyticity:} $M^{(1)}(z)$ is analytic in $\mathbb{C}\setminus\Sigma^{(1)}$, where $\Sigma^{(1)}$, shown in Fig.~\ref{fig:Hirota3}, is defined by
	\begin{equation}
		\Sigma^{(1)}=(-\infty,z_1]\cup(\cup_{j=0}^n[E_j,\widehat{E}_j])\cup\Sigma_1\cup\Sigma_1^*\cup\Sigma_2\cup\Sigma_2^*.
	\end{equation}
	
	\item \textbf{Jump condition:} $M^{(1)}(z)$ satisfies the jump relation $M^{(1)}_+(z)=M^{(1)}_-(z)J_{M}^{(1)}(z)$ for $z\in\Sigma^{(1)}$, where the jump matrix is given by
	\begin{equation} \label{eq:M1_jump}
		J_{M}^{(1)}(z)=\begin{cases}
			-ie^{i(f_0x+(\alpha g_0+\beta h_0)t-(B_j^fx+(\alpha B_j^g+\beta B_j^h)t+\phi_j-\delta_j)/2)\hat{\sigma}_3}\sigma_1,\v\\
			\begin{pmatrix}1&0 \v\\-\overline{r_1(z)}e^{-2it\theta(z)-2\delta(z)}&1\end{pmatrix},&z\in\Sigma_1,\v\\
			\begin{pmatrix}1&r_1(z)e^{2it\theta(z)+2\delta(z)}\\0&1\end{pmatrix},&z\in\Sigma_1^*,\\[2.5ex]
			\begin{pmatrix}1&\dfrac{r_1(z)e^{2it\theta(z)+2\delta(z)}}{1-|r_1(z)|^2}\v\\0&1\end{pmatrix},&z\in\Sigma_2,v\\\
			\begin{pmatrix}1&0 \v\\ \dfrac{-\overline{r_1(z)}e^{-2it\theta(z)-2\delta(z)}}{1-|r_1(z)|^2}&1\end{pmatrix},&z\in\Sigma_2^*,\v\\
			\begin{pmatrix}1-|r_1(z)|^2&r_1(z)e^{2it\theta(z)+2\delta(z)} \v\\ -\overline{r_1(z)}e^{-2it\theta(z)-2\delta(z)}&1\end{pmatrix},&z\in(\widehat{E}_{j_1},z_1),
		\end{cases}
	\end{equation}
	with $\phi_0=\delta_0=0$ and $j=0,\dots,n$.
	
	\item \textbf{Normalization at infinity:} As $z\to\infty$, we have $M^{(1)}(z)=I+\mathcal{O}(z^{-1})$.
	
	\item \textbf{Local behavior:} For $p\in\{E_j,\widehat{E}_j\}_{j=0}^n\setminus\{\widehat{E}_{j_1}\}$,\,\, $
		M^{(1)}(z)=\mathcal{O}((z-p)^{-1/4}),\quad z\to p.$
\end{enumerate}
\end{RH}

\subsection{Global parametrix}
We now construct the global parametrix, which captures the leading asymptotic behavior of the Riemann--Hilbert problem. We seek a global parametrix $M^{(glo)}$ solving the model Riemann--Hilbert problem obtained by neglecting the exponentially small jumps on the lens contours. More precisely, we consider the following problem:

\begin{RH} $M^{(glo)}(z)$ satisfies the following Riemann--Hilbert problem:

\begin{enumerate}[label=(\textbf{$M^{(glo)}$\arabic*}), leftmargin=*]
	\item \textbf{Analyticity:} $M^{(glo)}(z)$ is analytic in $\mathbb{C}\setminus(\cup_{j=0}^n[E_j,\widehat{E}_j])$.
	
	\item \textbf{Jump condition:}  $M^{(glo)}_+(z)=M^{(glo)}_-(z)J_{M}^{(glo)}(z)$ for $z\in\cup_{j=0}^n(E_j,\widehat{E}_j)$, where the jump matrix is given by
	\begin{equation} \label{eq:M^{(glo)}_jump}
		J_{M}^{(glo)}(z)=
		-ie^{i(f_0x+(\alpha g_0+\beta h_0)t-(B_j^fx+(\alpha B_j^g+\beta B_j^h)t+\phi_j-\delta_j)/2)\hat{\sigma}_3}\sigma_1,\quad z\in(E_j,\widehat{E}_j),
	\end{equation}
	with $\phi_0=\delta_0=0$ and $j=0,\dots,n$.
	
	\item \textbf{Normalization at infinity:} As $z\to\infty$,  $M^{(glo)}(z)=I+\frac{M_1^{(glo)}}{z}+\mathcal{O}(z^{-2})$.
	
	\item \textbf{Local behavior:} For $p\in\{E_j,\widehat{E}_j\}_{j=0}^n$,
	$		M^{(glo)}(z)=\mathcal{O}((z-p)^{-1/4}),\quad z\to p.$
	\end{enumerate}
 \end{RH}

As in the Riemann--Hilbert problem for $\mu^{(AG)}$, the above problem can be solved explicitly using the planar matrix Baker--Akhiezer function. More precisely, we have
\begin{equation}\label{M_glo}
	M^{(glo)}(z)=e^{-i(f_0x+(\alpha g_0+\beta h_0)t)\sigma_3}D^{(glo)}(\infty)^{-1}D^{(glo)}(z)e^{i(f_0x+(\alpha g_0+\beta h_0)t)\sigma_3},
\end{equation}
where $D^{(glo)}(z)=\left(D_{ij}^{(glo)}(z)\right)_{i,j=1,2}$ with
\begin{equation}
	D_{jj}^{(glo)}(z)=\frac{1}{2}(\kappa(z)+\kappa(z)^{-1})W_{jj}(z;x,t,\boldsymbol{\phi}-\boldsymbol{\delta}),\quad j=1,2,
\end{equation}
\begin{equation}
	D_{ij}^{(glo)}(z)=\frac{1}{2}(\kappa(z)-\kappa(z)^{-1})W_{ij}(z;x,t,\boldsymbol{\phi}-\boldsymbol{\delta}),\quad
   i=1,2;\, j=3-i.
\end{equation}

Consequently, we obtain
\begin{equation}\label{M_glo_12}
	q^{(AG)}(x,t;\boldsymbol{E},\boldsymbol{\widehat{E}},\boldsymbol{\phi}-\boldsymbol{\delta})=2i\left(M_1^{(glo)}\right)_{12}.
\end{equation}

\subsection{Local parametrix}
The global parametrix $M^{(glo)}(z)$ constructed in the previous subsection gives an accurate approximation to $M^{(1)}(z)$ away from the endpoint $\widehat{E}_{j_1}$. At the endpoint $\widehat{E}_{j_1}$, however, this approximation fails because the jump matrices for $M^{(1)}(z)$ do not converge uniformly to the identity and may exhibit singular behavior. To resolve this issue, we construct a local parametrix in a small neighborhood of this point.

We define the small open disk $U_{1}=\{z\in\mathbb{C}: |z-\widehat{E}_{j_1}|<\delta_1\}$ centered at $\widehat{E}_{j_1}$, with fixed radius $\delta_1>0$ chosen sufficiently small. In particular, we may choose
\begin{equation}
	\delta_1=\min\left\{\frac{1}{2}(\widehat{E}_{j_1}-z_{j_1}),\frac{1}{2}(E_{j_1+1}-\widehat{E}_{j_1}),2(z_1-\widehat{E}_{j_1})t^\varrho\right\},\quad\varrho\in(1/3,2/3).
\end{equation}

%We seek a matrix-valued function $M^{(loc)}(z)$ satisfying the following Riemann--Hilbert problem:

\begin{RH} $M^{(loc)}(z)$ satisfying the following Riemann--Hilbert problem:

\begin{enumerate}[label=(\textbf{$M^{(loc)}$\arabic*}), leftmargin=*]
	\item \textbf{Analyticity:} $M^{(loc)}(z)$ is analytic in $U_1\setminus\Sigma^{(1)}$.
	
	\item \textbf{Jump condition:} $M^{(loc)}_+(z)=M^{(loc)}_-(z)J_{M}^{(1)}(z)$ for $z\in\Sigma^{(1)}\cap U_1$.
	
	\item \textbf{Matching condition:} As $t\to\infty$, $M^{(loc)}(z)$ matches $M^{(glo)}(z)$ on the boundary $\partial U_1$.
	
	\item \textbf{Local behavior:} At $\widehat{E}_{j_1}$, $
		M^{(loc)}(z)=\mathcal{O}((z-\widehat{E}_{j_1})^{-1/4}),\quad z\to \widehat{E}_{j_1}.$
	
\end{enumerate}
\end{RH}

To construct the solution, we need a suitable conformal map that captures the local behavior of the phase function near the endpoint. For $\theta$, we have the following local expansion:
\begin{equation}
	\theta(z)=\theta^{(0,\hat{j}_1)}+(\xi-\widehat{\xi}_{j_1})\theta^{(1,\hat{j}_1)}(z-\widehat{E}_{j_1})^{1/2}+\frac{2}{3}\theta^{(3,\hat{j}_1)}(z-\widehat{E}_{j_1})^{3/2}+O\left((z-\widehat{E}_{j_1})^{5/2}\right),
\end{equation}
as $z\to\widehat{E}_{j_1}$, for fixed $\xi$, where
\begin{equation}
	\theta^{(0,\hat{j}_1)}=\theta(\widehat{E}_{j_1};\xi)=f_0\xi+\alpha g_0+\beta h_0-\frac{1}{2}(B_{j_1}^f\xi+\alpha B_{j_1}^g+\beta B_{j_1}^h),
\end{equation}
\begin{equation}
	\theta^{(1,\hat{j}_1)}=-\frac{2\prod_{j=0}^n(\widehat{E}_{j_1}-z_j^f)}{\hat{w}^{(j_1)}(\widehat{E}_{j_1})},
\end{equation}
\begin{equation}
	\theta^{(3,\hat{j}_1)}=\frac{1}{\hat{w}^{(j_1)}(\widehat{E}_{j_1})^2}\left[F(\widehat{E}_{j_1};\xi)(\hat{w}^{(j_1)})^{\prime}(\widehat{E}_{j_1})-\partial_zF(\widehat{E}_{j_1};\xi)\hat{w}^{(j_1)}(\widehat{E}_{j_1})\right],
\end{equation}
with
\begin{equation}
	w^{(i)}(z)=\left[-\prod_{k=0}^n(z-\widehat{E}_k)\cdot\prod_{k=0,...,n,k\neq i}(z-E_k)\right]^{\frac{1}{2}},
\end{equation}
and
\begin{equation}
	\hat{w}^{(i)}(z)=\left[\prod_{k=0}^n(z-E_k)\cdot\prod_{k=0,...,n,k\neq i}(z-\widehat{E}_k)\right]^{\frac{1}{2}}.
\end{equation}

This local expansion motivates the definition
\begin{equation}
	\zeta(z)=\zeta(z;\xi)=\left(\frac{3it}{2}(\theta(\widehat{E}_{j_1})+(\xi-\widehat{\xi}_{j_1})\theta^{(1,\hat{j}_1)}(z-\widehat{E}_{j_1})^{1/2}-\theta(z))\right)^{2/3},\quad z\in U_1,
\end{equation}
which is a one-to-one conformal map in $U_1$ with respect to $z$. Moreover, it is readily seen that
\begin{equation}
	\zeta(\widehat{E}_{j_1})=0,\quad\zeta^{\prime}(\widehat{E}_{j_1})=-|\theta^{(3,\hat{j}_1)}(\xi)|^{2/3}t^{2/3}<0.
\end{equation}

\subsection{Construction of a model Riemann--Hilbert problem}

In this subsection, we explicitly construct the matrix function $M^{(loc)}(z)$ using the Painlev\'{e} XXXIV function. The construction is carried out in three steps: first, we solve a standard model Riemann--Hilbert problem in the auxiliary $\zeta$-plane; next, we establish the error estimate; finally, we construct $M^{(loc)}(z)$.

We use the standard Painlev\'{e} XXXIV parametrix $\Psi(\zeta)$, as described in \cite{Fan2026}.
\begin{RH} $\Psi(\zeta)$ satifies the following RH problem:

\begin{enumerate}[label=(\textbf{$\Psi$\arabic*}), leftmargin=*]

	\item \textbf{Analyticity:} $\Psi(\zeta)$ is analytic in $\mathbb{C}\setminus \Sigma^{(loc,\zeta)}$ and continuous up to the boundary. The jump contour $\Sigma^{(loc,\zeta)}$ is defined by
	\begin{equation}
		\begin{aligned}
			\Sigma^{(loc,\zeta)} &=(\zeta(z_1),+\infty)\cup\Sigma_1^{(loc)}\cup\Sigma_1^{(loc)*},\v\\ \Sigma_1^{(loc)} &=(\zeta(\Sigma_1^*\cap U))\cup\{\zeta(\Sigma_1^*\cap\partial U)+e^{i(\pi-\varphi)}\mathbb{R}^+\}.
		\end{aligned}
	\end{equation}
	
	\item \textbf{Jump condition:} $\Psi(\zeta)$ satisfies the jump relation $\Psi_+(\zeta)=\Psi_-(\zeta)J_M^{(loc,1)}(\zeta)$, where the jump matrices are given by
	\begin{equation}
		J_M^{(loc,1)}(\zeta)=e^{it\theta(\widehat{E}_{j_1})\hat{\sigma}_3}
		\begin{cases}
			ie^{-i(\phi_{j_{1}}-\delta_{j_{1}})\hat{\sigma}_{3}/2}\sigma_1
			, & \zeta\in\mathbb{R}^{+}, \v\\
			\begin{pmatrix}
				1 & e^{-i\phi_{j_{1}}+2\hat{\theta}(\zeta)+i\delta_{j_{1}}} \\
				0 & 1
			\end{pmatrix}, & \zeta\in\Sigma_{1}^{(loc)}, \v\\
			\begin{pmatrix}
				1 & 0 \\
				e^{i\phi_{j_{1}}+2\hat{\theta}(\zeta)-i\delta_{j_{1}}} & 1
			\end{pmatrix}, & \zeta\in\Sigma_{1}^{(loc)*},\v \\
			\begin{pmatrix}
				1 & 0 \\
				e^{i\phi_{j_{1}}+2\hat{\theta}(\zeta)-i\delta_{j_{1}}} & 1
			\end{pmatrix}
			\begin{pmatrix}
				1 & e^{-i\phi_{j_{1}}+2\hat{\theta}(\zeta)+i\delta_{j_{1}}} \\
				0 & 1
			\end{pmatrix}, & \zeta\in(\zeta(z_{1}),0),
		\end{cases}
	\end{equation}
	where
	\begin{equation}
		\hat{\theta}(\zeta)=s\zeta^{1/2}+\frac{2}{3}\zeta^{3/2},
	\qquad
		s=-\frac{\theta^{(1,\hat{j}_1)}}{|\theta^{(3,\hat{j}_1)}(\xi)|^{1/3}}(\xi-\widehat{\xi}_{j_1})t^{2/3}\in\mathbb{R}.
	\end{equation}
	
	\item \textbf{Asymptotics at infinity:} As $\zeta\to\infty$,
	\begin{equation}
		\Psi(\zeta)=\left(I+\mathcal{O}\left(\zeta^{-1}\right)\right)\frac{i\zeta^{-\frac{1}{4}\sigma_3}}{\sqrt{2}}
		 \sigma_1G_1(\zeta),
	\end{equation}
	where
	\begin{equation}
		G_1(\zeta)=
		\begin{cases}
			\begin{pmatrix}
				0 & -1 \\
				1 & 0
			\end{pmatrix}e^{\pi i\sigma_3/4}e^{-i(t\theta(\widehat{E}_{j_1})-\phi_{j_1}+\delta_{j_1})\sigma_3/2}, & \zeta\in\mathbb{C}^+, \v \\
			e^{\pi i\sigma_3/4}e^{-i(t\theta(\widehat{E}_{j_1})-\phi_{j_1}+\delta_{j_1})\sigma_3/2}, & \zeta\in\mathbb{C}^-.
		\end{cases}
	\end{equation}
	
	\item \textbf{Local behavior:} As $\zeta\to0$,  $\Psi(\zeta)=\mathcal{O}(\zeta^{-1/4})$.
\end{enumerate}
\end{RH}

This model problem admits an explicit solution in terms of the Painlev\'{e} XXXIV function $M^{(P_{34})}(\zeta;s,\alpha,\omega)$. To this end, let $\alpha=-1/4$ and $\omega=0$, and define
\begin{equation}
	\Psi(\zeta)=
	\begin{pmatrix}
		1 & 0 \\
		i\nu(s) & 1
	\end{pmatrix}M^{(P_{34})}(\zeta;s,-1/4,0)e^{\hat{\theta}(\zeta)\sigma_3}G_1(\zeta)G_2(\zeta),
\end{equation}
where
\begin{equation}
	\begin{aligned}
		G_2(\zeta)=e^{it\theta(\widehat{E}_{j_1})\hat{\sigma}_3}
		\begin{cases}
			\begin{pmatrix}
				1 & e^{-i\phi_{j_1}+2\hat{\theta}(\zeta)+i\delta_{j_1}} \\
				0 & 1
			\end{pmatrix}, & \zeta\in\Sigma_1^{(loc)}, \v\\
			\begin{pmatrix}
				1 & 0 \\
				e^{i\phi_{j_1}+2\hat{\theta}(\zeta)-i\delta_{j_1}} & 1
			\end{pmatrix}, & \zeta\in\Sigma_1^{(loc)*}, \v\\
			I, & \text{elsewhere,}
		\end{cases}
	\end{aligned}
\end{equation}
with
\begin{equation}\label{a}
	\nu(s)=\int_{-\infty}^s\left(P_{34}(\omega)+\frac{\omega}{2}\right)\mathrm{d}\omega.
\end{equation}
Here $P_{34}(s)$ is the unique solution of the Painlev\'{e}-XXXIV equation \cite{Fokas2006}
\begin{equation}
	P_{34}^{\prime\prime}(s)=4P_{34}(s)^2+2sP_{34}(s)+\frac{4P_{34}^{\prime}(s)^2-1}{8P_{34}(s)}.
\end{equation}

Note that, as $t\to\infty$, $J_M^{(1)}(\zeta)$ is well approximated by $J_M^{(loc,1)}(\zeta)$ for $\zeta\in\Sigma_1^{(loc)}\cap\zeta(U)$. It is therefore natural to expect that $M^{(loc)}$ is well approximated by $\Psi$ for large $t$. We begin by introducing a Riemann--Hilbert problem whose jump matrix is the same as that of $M^{(loc)}$ and whose large-$\zeta$ asymptotics agree with those of $\Psi$.

\begin{RH} ${M}^{(loc,0)}(\zeta)$ satisfies the following RH problem:

\begin{enumerate}[label=(\textbf{${M}^{(loc,0)}$\arabic*}), leftmargin=*]
	\item \textbf{Analyticity:} ${M}^{(loc,0)}(\zeta)$ is analytic in $\mathbb{C}\setminus \Sigma^{(loc,\zeta)}$.
	
	\item \textbf{Jump condition:}  ${M}^{(loc,0)}_+(\zeta)={M}^{(loc,0)}_-(\zeta)J_M^{(1)}(\zeta),\quad \zeta\in \Sigma^{(loc,\zeta)}$.
	
	\item \textbf{Asymptotics at infinity:} As $\zeta\to\infty$,
	\begin{equation}
		{M}^{(loc,0)}(\zeta)=\frac{1}{\sqrt{2}}\left(I+\mathcal{O}\left(\zeta^{-1}\right)\right)\zeta^{-\frac{1}{4}\sigma_3}
		\begin{pmatrix}
			1 & i \\
			i & 1
		\end{pmatrix}G_1(\zeta),
	\end{equation}
	
	\item \textbf{Local behavior:} As $\zeta\to0$, ${M}^{(loc,0)}(\zeta)=\mathcal{O}(\zeta^{-1/4})$.
\end{enumerate}
 \end{RH}

Set $\Xi(\zeta)=M^{(loc,0)}(\zeta)\Psi(\zeta)^{-1}$. Following \cite{Fan2026}, one obtains that, as $t\to\infty$, $\Xi(\zeta)$ exists uniquely and satisfies
\begin{equation}
	\Xi(\zeta)=I+\mathcal{O}(t^{-1/3}).
\end{equation}

Hence the asymptotic expansion of $M^{(loc,0)}(\zeta)$ is
\begin{equation}
	M^{(loc,0)}(\zeta)=\frac{1}{\sqrt{2}}\left(I+\frac{M_1^{(loc,0)}}{\zeta}+O\left(\zeta^{-2}\right)\right)
\zeta^{-\frac{1}{4}\sigma_3}
	\begin{pmatrix}
		1 & i \\
		i & 1
	\end{pmatrix}G_1(\zeta),
\end{equation}
where
\begin{equation}
	(M_1^{(loc,0)})_{12}=i\nu(s)+O(t^{-1/3}).
\end{equation}

We now construct $M^{(loc)}(z)$. With the aid of $M^{(loc,0)}(\zeta)$, we finally define
\begin{equation}
	M^{(loc)}(z)=H_1(z)t^{\sigma_3/6}M^{(loc,0)}(\zeta(z)),\quad z\in U_1,
\end{equation}
where
\begin{equation}
	H_1(z)=\frac{1}{\sqrt{2}}M^{(glo)}(z)G_1(\zeta(z))^{-1}\begin{pmatrix}
		1 & -i \\
		-i & 1
	\end{pmatrix}\left((\widehat{E}_{j_1}-z)|\theta^{(3,\hat{j}_1)}(\widehat{\xi}_{j_1})|^{2/3}\right)^{\sigma_3/4}.
\end{equation}
The function $H_1(z)$ is analytic in $U_1$; see \cite{Fan2026}. Finally, using $M^{(glo)}(z)$ and $M^{(loc)}(z)$, we obtain
\begin{equation}
	M^{(loc)}(z)M^{(glo)}(z)^{-1}=I-\frac{t^{1/3}}{\zeta(z)}H_1(z)
	\begin{pmatrix}
		0 & i\nu(s) \\
		0 & 0
	\end{pmatrix}H_1(z)^{-1}+\mathcal{O}(t^{1/3-2\varrho}),
\end{equation}
as $t\to\infty$, for $z\in\partial U_1$.

\subsection{The small-norm Riemann--Hilbert problem}
The final transformation involves the global parametrix $M^{(glo)}(z)$ and the local parametrix $M^{(loc)}(z)$. We define the matrix-valued function $E(z)$ by
\begin{equation}
	E(z)=
	\begin{cases}
		M^{(1)}(z)M^{(glo)}(z)^{-1},\qquad z\in\mathbb{C}\setminus U_1, \\
		M^{(1)}(z)M^{(loc)}(z)^{-1},\qquad z\in U_1,
	\end{cases}
\end{equation}
which satisfies the following Riemann--Hilbert problem.

\begin{figure}[!t]
	\centering
	\includegraphics[scale=0.25]{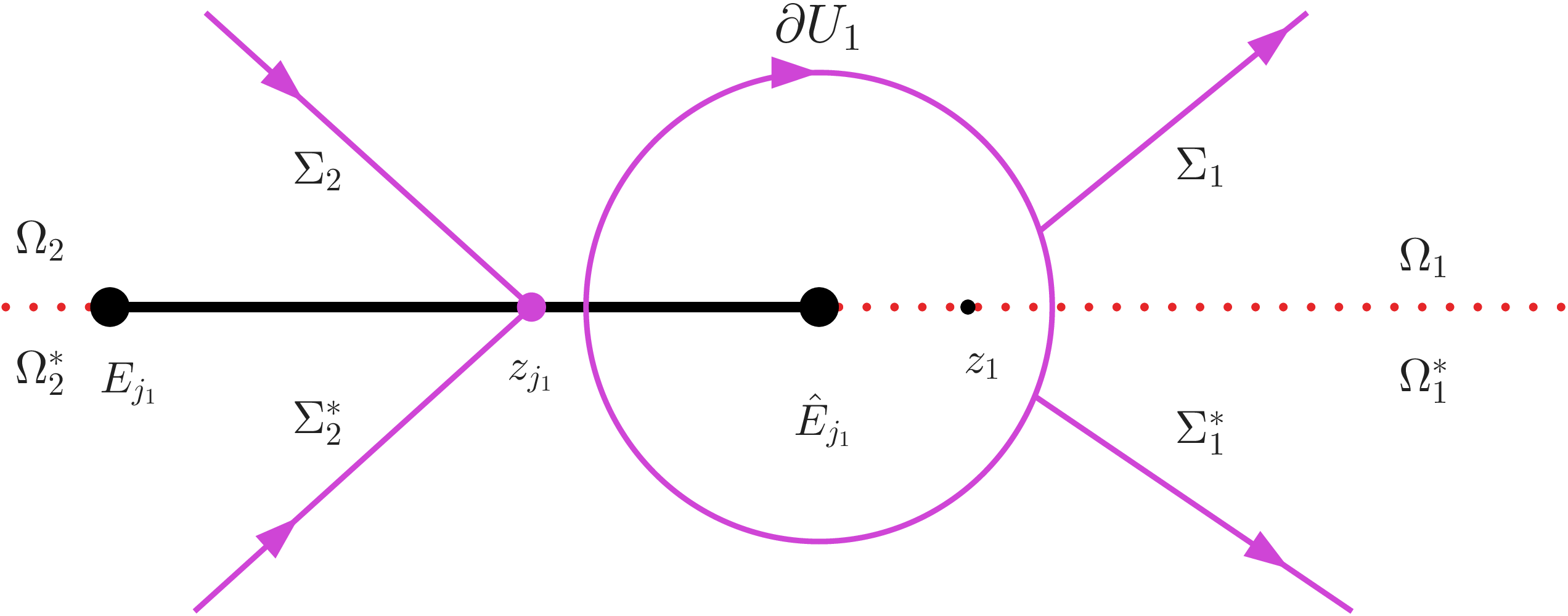}
	\caption{The contours $\Sigma^{(E)}$.}
	\label{fig:Hirota4}
\end{figure}

\begin{RH} $E(z)$ satisfies the following Riemann--Hilbert problem:

\begin{enumerate}[label=(\textbf{$E$\arabic*}), leftmargin=*]
	
\item \textbf{Analyticity:} $E(z)$ is analytic in $\mathbb{C}\setminus \Sigma^{(E)}$, where $
			\Sigma^{(E)}=\partial U_1\cup\Sigma_2\cup\Sigma_2^*\cup\Sigma_1\cup\Sigma_1^*\setminus U_1$ (see Fig.~\ref{fig:Hirota4}).

	\item \textbf{Jump condition:} $E_+(z)=E_-(z)J_{E}(z)$, where the jump matrix is given by
	\begin{equation}
		J_{E}(z)=\left\{
		\begin{array}{ll}
			M^{(glo)}(z)J_M^{(1)}(z)M^{(glo)}(z)^{-1}, & z\in\Sigma^{(E)}\setminus\partial U_1, \v\\
			M^{(loc)}(z)M^{(glo)}(z)^{-1}, & z\in\partial U_1.
		\end{array}\right.
	\end{equation}
	
	\item \textbf{Asymptotics at infinity:} As $z\to\infty$,  $E(z)=I+\frac{E_1}{z}+\mathcal{O}(z^{-2})$.
	
	\item \textbf{Local behavior:} As $z\to\widehat{E}_{j_1}$, $E(z)=\mathcal{O}((z-\widehat{E}_{j_1})^{-1/2})$.
\end{enumerate}
\end{RH}

A direct calculation shows that
\begin{equation}
	\parallel J_E(z)-I\parallel_{L^p}=
	\begin{cases}
		\mathcal{O}\left(e^{-ct^{\varrho/2}}\right), & z\in\Sigma^{(E)}\setminus\partial U_1, \\
		\mathcal{O}(t^{-\kappa_p}), & z\in\partial U_1 ,
	\end{cases}
\end{equation}
for some positive constant $c$, with $\kappa_\infty=\varrho-1/3$ and $\kappa_2=\varrho/2$; see \cite{Fan2026}. Since the jump matrix is uniformly close to the identity, the standard small-norm Riemann--Hilbert theory (see, e.g., \cite{Deift1993}) implies that the Riemann--Hilbert problem for $E$ has a unique solution for sufficiently large positive $t$. Moreover,
\begin{equation}
	E(z)=I+\frac{1}{2\pi i}\int_{\Sigma^{(E)}}\frac{\varpi(\lambda)(J_E(\lambda)-I)}{\lambda-z}\mathrm{d}\lambda,
\end{equation}
where $\varpi\in I+L^2(\Sigma^{(E)})$ is the unique solution of the Fredholm-type equation
\begin{equation}
	\varpi=I+\mathcal{C}_E\varpi,
\end{equation}
and $\mathcal{C}_E{:}L^2(\Sigma^{(E)})\to L^2(\Sigma^{(E)})$ is the integral operator defined by $\mathcal{C}_E(f)(z)=\mathcal{C}_-\left(f(J_E(z)-I)\right)$, with $\mathcal{C}_-$ being the Cauchy projection operator on $\Sigma^{\large(E)}$ defined by
\begin{equation}
	\mathcal{C}_{\pm}f(z)=\lim_{z^{\prime}\to z\in\Sigma^{\large(E)}}\frac{1}{2\pi i}\int_{\Sigma^{\large(E)}}\frac{f(\zeta)}{\zeta-z^{\prime}}\mathrm{d}\zeta.
\end{equation}
It is readily seen that
\begin{equation}
	E_1=-\frac{1}{2\pi i}\int_{\Sigma^{(E)}}\varpi(\lambda)(J_E(\lambda)-I)\mathrm{d}\lambda.
\end{equation}
As in \cite{Fan2026}, we obtain
\begin{equation}\label{E_1}
	E_1=-\frac{t^{-1/3}}{|\theta^{(3,\hat{j}_1)}(\widehat{\xi}_{j_1})|^{2/3}}H_1(\widehat{E}_{j_1})
	\begin{pmatrix}
		0 & ia \\
		0 & 0
	\end{pmatrix}H_1(\widehat{E}_{j_1})^{-1}+\mathcal{O}(t^{-\varrho}).
\end{equation}

\section{Long-time asymptotics in the transition region II}\label{Region-II}

In this section, we carry out the asymptotic analysis of the Riemann--Hilbert problem for $N(x,t;z)$ in order to derive the asymptotics of $q$ in transition region II. According to Definition~2.1, we may choose $C$ so that, throughout this section, only one of the saddle points $z_2$ is involved in the transition regime $0\le(\xi-\xi_{j_2})t^{2/3}\le C $ for some fixed $j_2\in\{0,\ldots,n\}$, since the analysis in the complementary half-region is analogous. For large $t$, one has $\xi_{j_2}<\xi<\widehat{\xi}_{j_2-1}$, and hence the saddle point satisfies $z_2\in[\widehat{E}_{j_2-1},E_{j_2})$.

In this and the next section, we use the same notation as before, with the meaning understood from the corresponding context.

Let $z_{j_2}\in(E_{j_2},\widehat{E}_{j_2})$ be fixed. We set
\begin{equation}
	\begin{aligned}
		&\Omega_3=\Omega_3(\xi)=\{z\in\mathbb{C}:\pi-\varphi_2\leq\arg(z-z_2)\leq\pi\},\quad \Sigma_3=\Sigma_3(\xi)=z_2+e^{i\varphi_2}\mathbb{R}^+,\\
		&\Omega_4=\Omega_4(\xi)=\{z\in\mathbb{C}:0\leq\arg(z-z_{j_2})\leq\varphi_2\},\quad
\Sigma_4=\Sigma_4(\xi)=z_{j_2}+e^{i(\pi-\varphi_2)}\mathbb{R}^+,
	\end{aligned}
\end{equation}
where $\varphi_{2}$ is chosen such that
\begin{equation}
	\operatorname{Im}\theta(z)
	\begin{cases}
		<0,\qquad z\in\Sigma_3\setminus\{z_2\}, \\
		>0,\qquad z\in\Sigma_4\setminus\{z_{j_2}\}.
	\end{cases}
\end{equation}

The first transformation for $N$ is defined by
\begin{equation}
	N^{(1)}(z)=e^{\delta(\infty)\sigma_3}N(z)G_N(z)e^{-\delta(z)\sigma_3},
\end{equation}
where
\begin{equation}
	G_N(z)=
	\begin{cases}
		\begin{pmatrix}
			1 & -\overline{r_2(z)}e^{2it\theta(z)} \\
			0 & 1
		\end{pmatrix},\quad z\in\Omega_3, \v\\
		\begin{pmatrix}
			1 & 0 \\
			-r_2(z)e^{-2it\theta(z)} & 1
		\end{pmatrix},\quad z\in\Omega_3^*, \v\\
		\begin{pmatrix}
			1 & 0 \\
			\frac{r_2(z)e^{-2it\theta(z)}}{1-|r_2(z)|^2} & 1
		\end{pmatrix},\quad z\in\Omega_4, \v\\
		\begin{pmatrix}
			1 & \frac{\overline{r_2(z)}e^{2it\theta(z)}}{1-|r_2(z)|^2} \\
			0 & 1
		\end{pmatrix},\quad z\in\Omega_4^*, \\
		I, \quad \text{elsewhere,}
	\end{cases}
\end{equation}
Similarly, the function $\delta(z)$ is defined by
\begin{equation}
	\delta(z)=\frac{w(z)}{2\pi i}\left[\sum_{j=1}^n\delta_j\int_{E_j}^{\widehat{E}_j}\frac{i\mathrm{~d}s}{w_+(s)(s-z)}+\int_{(\widehat{E}_{j_2},+\infty)\setminus(\cup_{j=j_2}^{n}(E_j,\widehat{E}_j))}\frac{\log(1-|r_2(s)|^2)\mathrm{~d}s}{w(s)(s-z)}\right],
\end{equation}
where the logarithm is taken on the principal branch, and the constants $\delta_j$, $j=1,\ldots,n$, are determined by the linear system
\begin{equation}\label{delta_2}
	\int_{(\widehat{E}_{j_2},+\infty)\setminus(\cup_{j=j_2}^{n}(E_j,\widehat{E}_j))}\frac{\log(1-|r_2(s)|^2)s^k\mathrm{d}s}{w(s)}+\sum_{j=1}^n\delta_j\int_{E_j}^{\widehat{E}_j}\frac{is^k\mathrm{d}s}{w_+(s)}=0,\quad k=0,\ldots,n-1.
\end{equation}

\begin{RH} $\delta(z)=-\overline{\delta(\bar{z})}$ satisfies the following Riemann--Hilbert problem:

\begin{enumerate}[label=(\textbf{$\delta$\arabic*}), leftmargin=*]
	\item \textbf{Analyticity:} $\delta(z)$ is analytic in $\mathbb{C}\setminus(\cup_{j=0}^n[E_j,\widehat{E}_j]\cup[z_2,+\infty))$.
	
	\item \textbf{Jump condition:} $\delta(z)$ satisfies the jump relation
	\begin{equation}
		\left\{\begin{aligned}
		&	\delta_-(z)  =\delta_+(z)+\log(1-|r_2(z)|^2),\quad &z\in(\widehat{E}_{j_2},+\infty)\setminus(\cup_{j=j_2}^{n}[E_j,\widehat{E}_j]), \\
		&	\delta_-(z)=-\delta_+(z)+i\delta_j,\quad &z\in(E_j,\widehat{E}_j),\quad j=1,\ldots,n.
		\end{aligned} \right.
	\end{equation}
	
	\item \textbf{Normalization at infinity:} As $z \to \infty$,
	\begin{equation}
		\delta(z)=\delta(\infty)+\frac{\delta^{(1)}}{z}+\mathcal{O}(z^{-2}),\quad z\to\infty,
	\end{equation}
	where
	\begin{equation}\label{delta_infty_2}
		\begin{aligned}
			\delta(\infty) =-\frac{1}{2\pi i}\left[\int_{(\widehat{E}_{j_2},+\infty)\setminus(\cup_{j=j_2}^{n}(E_j,\widehat{E}_j))}\frac{\log(1-|r_2(s)|^2)s^n \mathrm{d}s}{w(s)} +\sum_{j=1}^n\delta_j\int_{E_j}^{\widehat{E}_j}\frac{is^n\mathrm{d}s}{w_+(s)}\right],
		\end{aligned}
	\end{equation}
	and
	\begin{equation}
		\begin{aligned}
			\delta^{(1)} & =-\delta(\infty)\sum_{j=0}^n(E_j+\widehat{E}_j) \\
			& \quad -\frac{1}{2\pi i}\left[\int_{(\widehat{E}_{j_2},+\infty)\setminus(\cup_{j=j_2}^{n}(E_j,\widehat{E}_j))}\frac{\log(1-|r_2(s)|^2)s^{n+1} \mathrm{d}s}{w(s)}+\sum_{j=1}^n\delta_j\int_{E_j}^{\widehat{E}_j}\frac{is^{n+1}\mathrm{d}s}{w_+(s)}\right].
		\end{aligned}
	\end{equation}
	
	\item \textbf{Local behavior:} As $z\to p\in\{E_j\}_{j=j_2+1}^{n}\cup\{\widehat{E}_j\}_{j=j_2}^{n}$ from $\mathbb{C}^+$, we have
	\begin{equation}
		e^{\delta(z)}=\mathcal{O}((z-p)^{-1/2}),
	\end{equation}
	and
	\begin{equation}
		\delta(z)=\frac{i}{2}\delta_{j_2}+\mathcal{O}((z-E_{j_2})^{1/2}),\quad z\to E_{j_2} \text{~from~} \mathbb{C}\setminus(-\infty,E_{j_2}).
	\end{equation}
\end{enumerate}
\end{RH}

\begin{RH}  $N^{(1)}$ satisfies the following Riemann--Hilbert problem:

\begin{figure}[!t]
	\centering
	\includegraphics[scale=0.25]{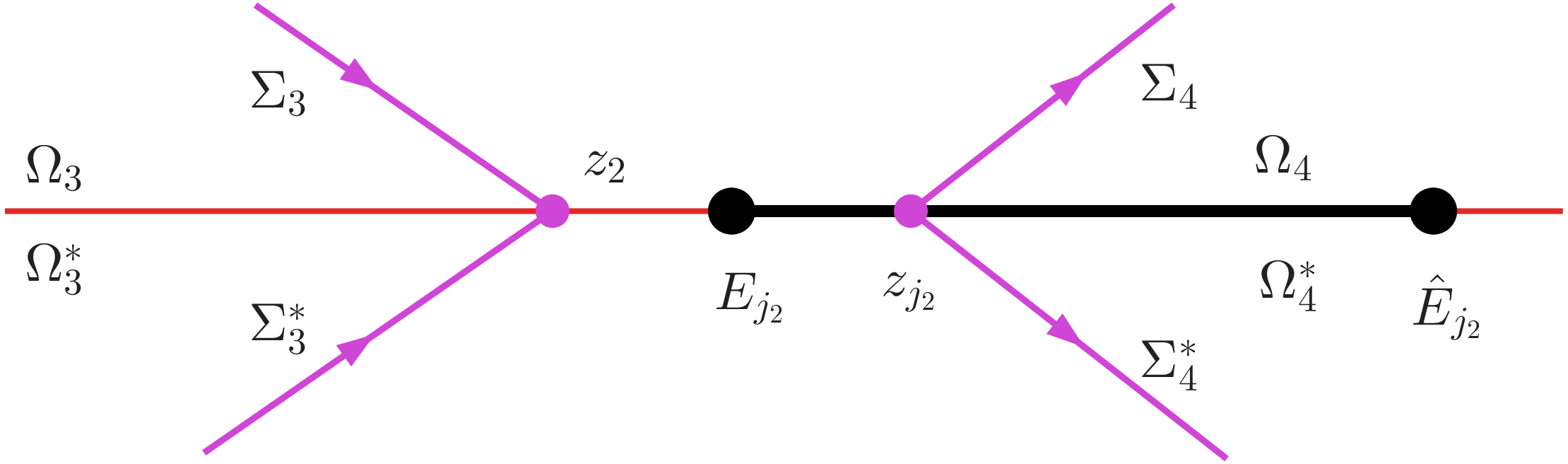}
	\caption{The contour $\Sigma^{(1)}$.}
	\label{fig:Hirota5}
\end{figure}

\begin{enumerate}[label=(\textbf{$N^{(1)}$\arabic*}), leftmargin=*]
	\item \textbf{Analyticity:} $N^{(1)}(z)$ is analytic in $\mathbb{C}\setminus\Sigma^{(1)}$, where $
		\Sigma^{(1)}=[z_2,E_{j_2}]\cup(\cup_{j=0}^n[E_j,\widehat{E}_j])\cup\Sigma_3\cup\Sigma_3^*\cup\Sigma_4\cup\Sigma_4^*$ (see Fig.~\ref{fig:Hirota5}).

	\item \textbf{Jump condition:} $N^{(1)}_+(z)=N^{(1)}_-(z)J_{N}^{(1)}(z)$ for $z\in\Sigma^{(1)}$, where the jump matrix is given by
	\begin{equation}
		J_{N}^{(1)}(z)=\begin{cases}
			-ie^{i(f_0x+(\alpha g_0+\beta h_0)t-(B_j^fx+(\alpha B_j^g+\beta B_j^h)t+\phi_j-\delta_j)/2)\hat{\sigma}_3}\sigma_1,&z\in(E_j,\widehat{E}_j), \v\\[2.5ex]
			\begin{pmatrix}1&\overline{r_2(z)}e^{2it\theta(z)+2\delta(z)}\\0&1\end{pmatrix},&z\in\Sigma_3, \v\\[2.5ex]
			\begin{pmatrix}1&0 \v\\ -r_2(z)e^{-2it\theta(z)-2\delta(z)}&1\end{pmatrix},&z\in\Sigma_3^*, \v\\[2.5ex]
			\begin{pmatrix}1&0 \v \\ \dfrac{-r_2(z)e^{-2it\theta(z)-2\delta(z)}}{1-|r_2(z)|^2}&1\end{pmatrix},&z\in\Sigma_4,\\[2.5ex]
			\begin{pmatrix}1&\dfrac{\overline{r_2(z)}e^{2it\theta(z)+2\delta(z)}}{1-|r_2(z)|^2}\\ 0&1\end{pmatrix},&z\in\Sigma_4^*,\v\\
			\begin{pmatrix}1&\overline{r_2(z)}e^{2it\theta(z)+2\delta(z)} \v\\ -r_2(z)e^{-2it\theta(z)-2\delta(z)}&1-|r_2(z)|^2\end{pmatrix},&z\in(z_2,E_{j_2}).
		\end{cases}
	\end{equation}
	
	\item \textbf{Normalization at infinity:} As $z\to\infty$,  $N^{(1)}(z)=I+\mathcal{O}(z^{-1})$.
	
	\item \textbf{Local behavior:} For $p\in\{E_j,\widehat{E}_j\}_{j=0}^n\setminus\{E_{j_2}\}$,
	 $
		N^{(1)}(z)=\mathcal{O}((z-p)^{-1/4}),\quad z\to p.$
	
\end{enumerate}
\end{RH}

Let
\begin{equation} \label{En}
	E(z)=
	\begin{cases}
		N^{(1)}(z)N^{(glo)}(z)^{-1}, & z\in\mathbb{C}\setminus U_2, \\
		N^{(1)}(z)N^{(loc)}(z)^{-1}, & z\in U_2,
	\end{cases}
\end{equation}
where $N^{(glo)}(z)$ and $N^{(loc)}(z)$ are the global and local parametrices for $N^{(1)}(z)$, respectively, and $U_{2}=\{z\in\mathbb{C}: |z-E_{j_2}|<\delta_2\}$ is centered at $E_{j_2}$, with fixed radius $\delta_2>0$ chosen sufficiently small. In particular, we may choose
\begin{equation}
	\delta_2=\min\left\{\frac{1}{2}(z_{j_2}-E_{j_2}),\frac{1}{2}(E_{j_2+1}-\widehat{E}_{j_2}),2(z_2-\widehat{E}_{j_2})t^\varrho\right\},\quad\varrho\in(1/3,2/3).
\end{equation}

Similarly to transition region I, $N^{(glo)}(z)$ satisfies the same global model Riemann--Hilbert problem, whose solution is similar to $M^{(glo)}(z)$ given by \eqref{M_glo} with
the parameters $\delta_j$'s replaced by (\ref{delta_2}).

 \begin{RH}  $N^{(loc)}(z)$ satisfies the following Riemann--Hilbert problem:
\begin{enumerate}[label=(\textbf{$N^{(loc)}$\arabic*}), leftmargin=*]
	\item \textbf{Analyticity:} $N^{(loc)}(z)$ is analytic in $U_2\setminus\Sigma^{(1)}$.
	
	\item \textbf{Jump condition:} $N^{(loc)}_+(z)=N^{(loc)}_-(z)J_{N}^{(1)}(z)$ for $z\in\Sigma^{(1)}\cap U_2$.
	
	\item \textbf{Matching condition:} As $t\to\infty$, $N^{(loc)}(z)$ matches $N^{(glo)}(z)$ on the boundary $\partial U_2$ of $U_2$.
	
	\item \textbf{Local behavior:} At $E_{j_2}$, $
		N^{(loc)}(z)=\mathcal{O}((z-E_{j_2})^{-1/4}),\quad z\to E_{j_2}.$
	
\end{enumerate}
\end{RH}

Similarly, to construct the solution, we need a suitable conformal map that captures the local behavior of the phase function near the endpoint. For $\theta$, we have the following local expansion:
\begin{equation}
	\theta(z)=\theta^{(0,j_2)}+(\xi-\xi_{j_2})\theta^{(1,j_2)}(E_{j_2}-z)^{1/2}+\frac{2}{3}\theta^{(3,j_2)}(E_{j_2}-z)^{3/2}+\mathcal{O}\left((E_{j_2}-z)^{5/2}\right),
\end{equation}
as $z\to E_{j_2}$, for fixed $\xi$, where
\begin{equation}
	\theta^{(0,j_2)}=f_0\xi+\alpha g_0+\beta h_0-\frac{1}{2}(B_{j_2}^f\xi+\alpha B_{j_2}^g+\beta B_{j_2}^h),
\end{equation}
\begin{equation}
	\theta^{(1,j_2)}=-\frac{2\prod_{j=0}^n(E_{j_2}-z_{j_2}^f)}{w^{(j_2)}(E_{j_2})},
\end{equation}
\begin{equation}
	\theta^{(3,j_2)}=\frac{-1}{w^{(j_2)}(E_{j_2})^2}\left[\partial_zF(E_{j_2};\xi)w^{(j_2)}(E_{j_2})-F(E_{j_2};\xi)(w^{(j_2)})^{\prime}(E_{j_2})\right].
\end{equation}

This local expansion motivates the definition
\begin{equation}
	\zeta(z)=\left(\frac{3it}{2}(\theta(E_{j_2})+(\xi-\xi_{j_2})\theta^{(1,j_2)}(E_{j_2}-z)^{1/2}-\theta(z))\right)^{2/3},\quad z\in U_2,
\end{equation}
which is a one-to-one conformal map in $U_2$ with respect to $z$. Moreover, it is readily seen that
\begin{equation}
	\zeta(E_{j_2})=0,\quad\zeta^{\prime}(E_{j_2})=-|\theta^{(3,j_2)}(\xi_{j_2})|^{2/3}t^{2/3}<0.
\end{equation}

As in transition region I, the local parametrix $N^{(loc)}(z)$ is given by
\begin{equation}
	N^{(loc)}(z)=\tilde{H}_1(z)t^{\sigma_3/6}
	\begin{pmatrix}
		1 & 0 \\
		i\nu(s) & 1
	\end{pmatrix}M^{(P_{34})}(\zeta(z);s,-1/4,0)e^{\hat{\theta}(\zeta(z))\sigma_3}G_1(\zeta(z))G_2(\zeta(z)),
\end{equation}
where
\begin{equation}
	G_1(\zeta)=
	\begin{cases}
		\begin{pmatrix}
			0 & -1 \\
			1 & 0
		\end{pmatrix}e^{-\pi i\sigma_3/4}e^{-i(f_0x+(\alpha g_0+\beta h_0)t-(B_{j_2}^fx+(\alpha B_{j_2}^g+\beta B_{j_2}^h)t+\delta_{j_2})\sigma_3/2}, & \zeta\in\mathbb{C}^-, \v \\
		e^{-\pi i\sigma_3/4}e^{-i(f_0x+(\alpha g_0+\beta h_0)t-(B_{j_2}^fx+(\alpha B_{j_2}^g+\beta B_{j_2}^h)t+\delta_{j_2})\sigma_3/2}, & \zeta\in\mathbb{C}^+,
	\end{cases}
\end{equation}
\begin{equation}
	\begin{aligned}
		G_2(\zeta)=e^{it\theta(E_{j_2})\hat{\sigma}_3}
		\begin{cases}
			\begin{pmatrix}
				1 & -\overline{r_2(E_{j_2})}e^{2\theta(\zeta)+i\delta_{j_2}} \\
				0 & 1
			\end{pmatrix}, & \zeta\in\Omega_3^{(loc)}, \v\\
			\begin{pmatrix}
				1 & 0 \\
				-r_2(E_{j_2})e^{2\theta(\zeta)-i\delta_{j_2}} & 1
			\end{pmatrix}, & \zeta\in\Omega_3^{(loc)*}, \v\\
			I, & \text{elsewhere,}
		\end{cases}
	\end{aligned}
\end{equation}
and
\begin{equation}
	\tilde{H}_1(z)=\frac{1}{\sqrt{2}}N^{(glo)}(z)G_1(\zeta(z))^{-1}\begin{pmatrix}
		1 & -i \\
		-i & 1
	\end{pmatrix}\left((z-E_{j_2})|\theta^{(3,j_2)}(\xi)|^{2/3}\right)^{\sigma_3/4}.
\end{equation}
Using $N^{(glo)}(z)$ and $N^{(loc)}(z)$, we obtain
\begin{equation}
	N^{(loc)}(z)N^{(glo)}(z)^{-1}=I-\frac{t^{1/3}}{\zeta(z)}\tilde{H}_1(z)
	\begin{pmatrix}
		0 & i\nu(s) \\
		0 & 0
	\end{pmatrix}\tilde{H}_1(z)^{-1}+\mathcal{O}(t^{1/3-2\varrho}),
\end{equation}
as $t\to\infty$, for $z\in\partial U_2$.

\begin{figure}[!t]
	\centering
	\includegraphics[scale=0.25]{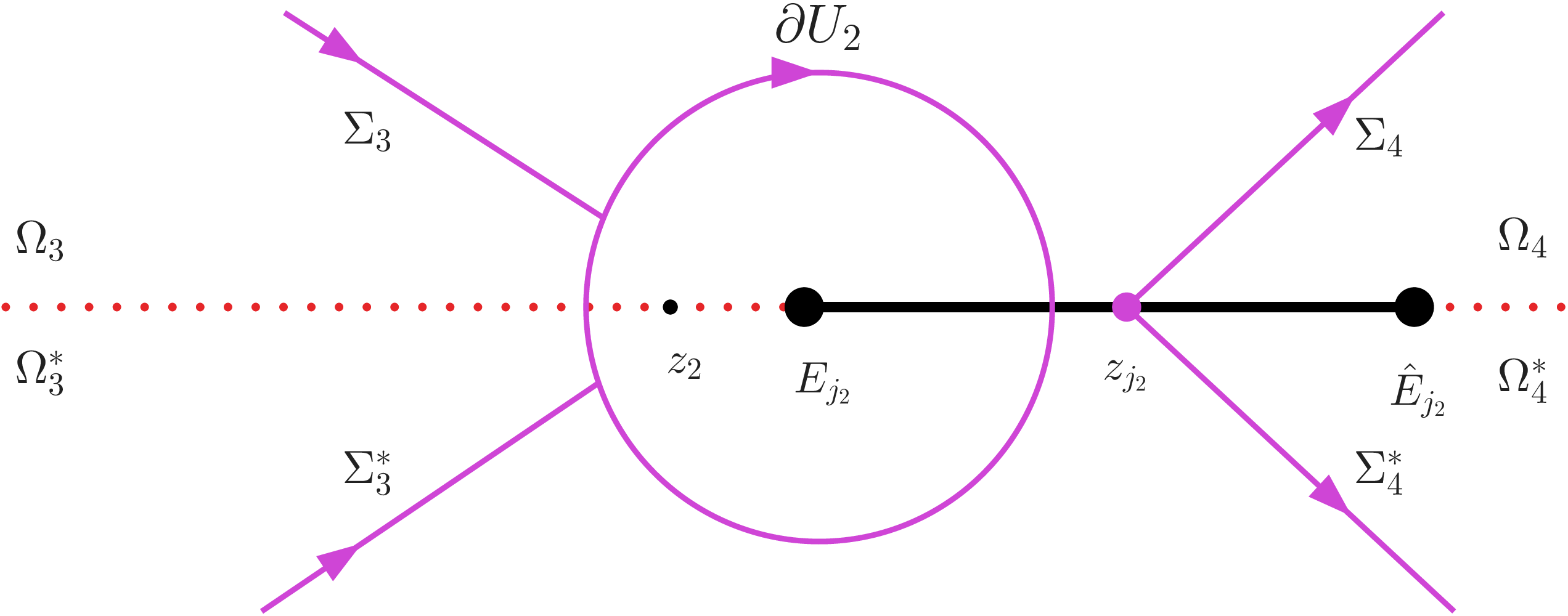}
	\caption{The contours $\Sigma^{(E)}$.}
	\label{fig:Hirota6}
\end{figure}

\begin{RH} $E(z)$ defined by (\ref{En}) satisfies the following Riemann--Hilbert problem:

\begin{enumerate}[label=(\textbf{$E$\arabic*}), leftmargin=*]
	\item \textbf{Analyticity:} $E(z)$ is analytic in $\mathbb{C}\setminus \Sigma^{(E)}$, where $
	\Sigma^{(E)}=\partial U_2\cup\Sigma_3\cup\Sigma_3^*\cup\Sigma_4\cup\Sigma_4^*\setminus U_2$ (see Fig.~\ref{fig:Hirota6}).

	\item \textbf{Jump condition:} $E_+(z)=E_-(z)J_{E}(z)$, where the jump matrix is given by
	\begin{equation}
		J_{E}(z)=\left\{
		\begin{array}{ll}
			N^{(glo)}(z)J_N^{(1)}(z)N^{(glo)}(z)^{-1}, & z\in\Sigma^{(E)}\setminus\partial U_2, \v\\
			N^{(loc)}(z)N^{(glo)}(z)^{-1}, & z\in\partial U_2.
		\end{array}\right.
	\end{equation}
	
	\item \textbf{Asymptotics at infinity:} As $z\to\infty$,  $E(z)=I+\frac{E_1}{z}+\mathcal{O}(z^{-2})$.
	
	\item \textbf{Local behavior:} As $z\to E_{j_2}$,  $E(z)=\mathcal{O}((z-E_{j_2})^{-1/2})$.
\end{enumerate}
\end{RH}

Similarly, by the standard small-norm Riemann--Hilbert theory, there exists a unique solution to the Riemann--Hilbert problem for $E$ for sufficiently large positive $t$. Moreover, we have
\begin{equation}
	E_1=\frac{t^{-1/3}}{|\theta^{(3,j_2)}(\xi_{j_2})|^{2/3}}\tilde{H}_1(E_{j_2})
	\begin{pmatrix}
		0 & i\nu(s) \\
		0 & 0
	\end{pmatrix}\tilde{H}_1(E_{j_2})^{-1}+\mathcal{O}(t^{-\varrho}),\quad t\to+\infty.
\end{equation}

\section{Long-time asymptotics in the Zakharov-Manakov region III and fast-decay region IV}\label{RegionIII-IV}

In this section, we carry out the asymptotic analysis of the Riemann--Hilbert problem for $M$ when $\xi$ belongs to the Zakharov--Manakov region, that is,
$\xi\in(-\infty,\widehat{\xi}_n)\cup_{j=1}^n(\xi_j,\widehat{\xi}_{j-1})\cup(\xi_0,+\infty)$, and to the fast-decay region, that is, $\xi\in\cup_{j=0}^n(\widehat{\xi}_j,\xi_j)$. In the Zakharov--Manakov region, the saddle points satisfy $z_i\in\mathbb{R}\setminus(\cup_{j=0}^n[E_j,\widehat{E}_j])$, $i=1,2$, whereas in the fast-decay region they satisfy $z_i\in\cup_{j=0}^n(E_j,\widehat{E}_j)$. We set
\begin{equation}
	\begin{aligned}
		&\Omega_1=\Omega_1(\xi)=\{z\in\mathbb{C}:0\leq\arg(z-z_1)\leq\varphi_1\},\,\, \Sigma_1=\Sigma_1(\xi)=z_1+e^{i\varphi_1}\mathbb{R}^+,\\
		&\Omega_2=\Omega_2(\xi)=\{z\in\mathbb{C}:\pi-\varphi_1\leq\arg(z-z_1)\leq\pi\},\,\, \Sigma_2=\Sigma_2(\xi)=z_1+e^{i(\pi-\varphi_1)}\mathbb{R}^+,\\
		&\Omega_3=\Omega_3(\xi)=\{z\in\mathbb{C}:0\leq\arg(z-z_2)\leq\varphi_2\},\,\, \Sigma_3=\Sigma_3(\xi)=z_2+e^{i\varphi_2}\mathbb{R}^+,\\
		&\Omega_4=\Omega_4(\xi)=\{z\in\mathbb{C}:\pi-\varphi_2\leq\arg(z-z_2)\leq\pi\},\,\, \Sigma_4=\Sigma_4(\xi)=z_2+e^{i(\pi-\varphi_2)}\mathbb{R}^+,
	\end{aligned}
\end{equation}
where $\varphi_{1}$ is chosen such that
\begin{equation}
	\operatorname{Im}\theta(z)
	\begin{cases}
		<0, \quad  z\in\Sigma_1\setminus\{z_1\}, \\
		>0,\quad z\in\Sigma_2\setminus\{z_1\},
	\end{cases}
\end{equation}
and $\varphi_{2}$ is chosen such that
\begin{equation}
	\operatorname{Im}\theta(z)
	\begin{cases}
		<0,\quad z\in\Sigma_3\setminus\{z_2\}, \\
		>0,\quad z\in\Sigma_4\setminus\{z_2\}.
	\end{cases}
\end{equation}

Similarly, the first transformation for $M$ is defined by
\begin{equation}
	M^{(1)}(z)=e^{\delta(\infty)\sigma_3}M(z)G(z)e^{-\delta(z)\sigma_3},
\end{equation}
where
\begin{equation}
	G(z)=
	\begin{cases}
		\begin{pmatrix}
			1 & 0 \\
			\overline{r_1(z)}e^{-2it\theta(z)} & 1
		\end{pmatrix},\quad z\in\Omega_1\cup\Omega_3, \v\\
		\begin{pmatrix}
			1 & r_1(z)e^{2it\theta(z)} \\
			0 & 1
		\end{pmatrix},\quad z\in\Omega_1^*\cup\Omega_3^*, \v\\
		\begin{pmatrix}
			1 & -\frac{r_1(z)e^{2it\theta(z)}}{1-|r_1(z)|^2} \\
			0 & 1
		\end{pmatrix},\quad z\in\Omega_2\cup\Omega_4, \v\\
		\begin{pmatrix}
			1 & 0 \\
			-\frac{\overline{r_1(z)}e^{-2it\theta(z)}}{1-|r_1(z)|^2} & 1
		\end{pmatrix},\quad z\in\Omega_2^*\cup\Omega_4^*, \v\\
		I, \quad \text{elsewhere.}
	\end{cases}
\end{equation}

The function $\delta(z)$ is defined by
\begin{equation}
	\delta(z)=\frac{w(z)}{2\pi i}\left[\sum_{j=1}^n\delta_j\int_{E_j}^{\widehat{E}_j}\frac{i\mathrm{~d}s}{w_+(s)(s-z)}-\int_{\cup_{i=1}^2(-\infty,z_i)\setminus(\cup_{j=0}^{n}(E_j,\widehat{E}_j))}\frac{\log(1-|r_1(s)|^2)\mathrm{~d}s}{w(s)(s-z)}\right],
\end{equation}
where the logarithm is taken on the principal branch, and the constants $\delta_j$, $j=1,\ldots,n$, are determined by the linear system
\begin{equation}\label{delta_3}
	\int_{\cup_{i=1}^2(-\infty,z_i)\setminus(\cup_{j=0}^{n}(E_j,\widehat{E}_j))}\frac{\log(1-|r_1(s)|^2)s^k\mathrm{d}s}{w(s)}-\sum_{j=1}^n\delta_j\int_{E_j}^{\widehat{E}_j}\frac{is^k\mathrm{d}s}{w_+(s)}=0,\quad k=0,\ldots,n-1.
\end{equation}

\begin{RH} $\delta(z)$ satisfies the following Riemann--Hilbert problem:

\begin{enumerate}[label=(\textbf{$\delta$\arabic*}), leftmargin=*]
	\item \textbf{Analyticity:} $\delta(z)$ is analytic in $\mathbb{C}\setminus(\cup_{j=0}^n[E_j,\widehat{E}_j]\cup(-\infty,z_1]\cup(-\infty,z_2])$.
	
	\item \textbf{Jump condition:} $\delta(z)$ satisfies the jump relation
	\begin{equation}
		\left\{\begin{aligned}
			\delta_-(z) & =\delta_+(z)+\log(1-|r_1(z)|^2),\quad &z\in\cup_{i=1}^2(-\infty,z_i)\setminus(\cup_{j=0}^{n}[E_j,\widehat{E}_j]), \\
			\delta_+(z) & +\delta_-(z)=i\delta_j,\quad &z\in(E_j,\widehat{E}_j),\quad j=1,\ldots,n.
		\end{aligned}\right.
	\end{equation}
	
	\item \textbf{Normalization at infinity:} As $z \to \infty$,
	$	\delta(z)=\delta(\infty)+\frac{\delta^{(1)}}{z}+\mathcal{O}(z^{-2}),$
		where
	\begin{equation}\label{delta_infty_3}
		\begin{aligned}
			\delta(\infty) =\frac{1}{2\pi i}\left[\int_{\cup_{i=1}^2(-\infty,z_i)\setminus(\cup_{j=0}^{n}(E_j,\widehat{E}_j))}\frac{\log(1-|r_1(s)|^2)s^n \mathrm{d}s}{w(s)} -\sum_{j=1}^n\delta_j\int_{E_j}^{\widehat{E}_j}\frac{is^n\mathrm{d}s}{w_+(s)}\right],
		\end{aligned}
	\end{equation}
	and
	\begin{equation}
		\begin{aligned}
			\delta^{(1)} & =\delta(\infty)\sum_{j=0}^n(E_j+\widehat{E}_j) \\
			&\qquad -\frac{1}{2\pi i}\left[\int_{\cup_{i=1}^2(-\infty,z_i)\setminus(\cup_{j=0}^{n}(E_j,\widehat{E}_j))}\frac{\log(1-|r_1(s)|^2)s^{n+1} \mathrm{d}s}{w(s)}-\sum_{j=1}^n\delta_j\int_{E_j}^{\widehat{E}_j}\frac{is^{n+1}\mathrm{d}s}{w_+(s)}\right].
		\end{aligned}
	\end{equation}
	
	\item \textbf{Local behavior:}  $		\delta(z)=\frac{i}{2}\delta_{i}+\mathcal{O}((z-z_{i})^{1/2}),\quad z\to z_{i} \text{~from~} \mathbb{C}\setminus(-\infty,z_{i}).$
	
\end{enumerate}
\end{RH}

\begin{figure}[!t]
	\centering
	\includegraphics[scale=0.35]{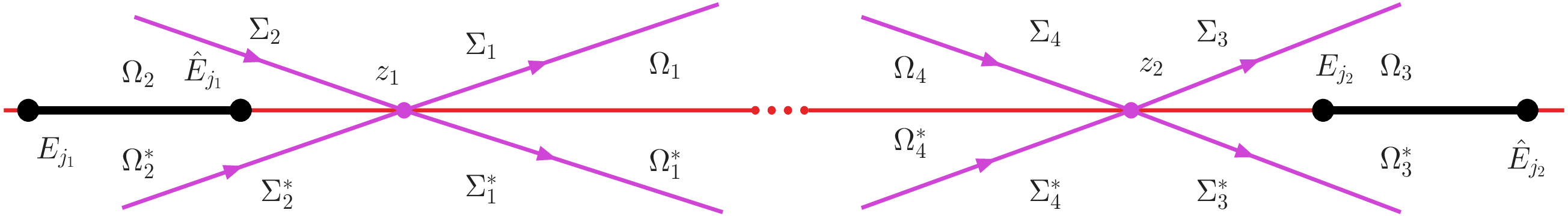}
	\caption{The contours $\Sigma^{(1)}$ in the Zakharov--Manakov region III.}
	\label{fig:Hirota71}
\end{figure}

\begin{figure}[!t]
	\centering
	\includegraphics[scale=0.35]{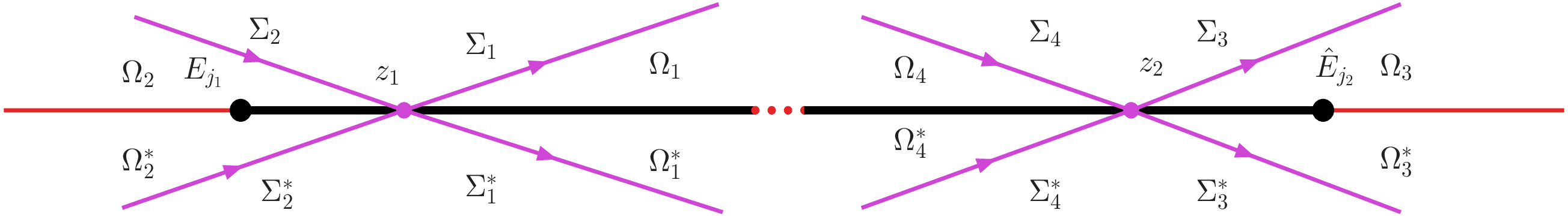}
	\caption{The contours $\Sigma^{(1)}$ in the fast-decay region IV.}
	\label{fig:Hirota72}
\end{figure}

\begin{RH} $M^{(1)}$ satisfies the following Riemann--Hilbert problem:

\begin{enumerate}[label=(\textbf{$M^{(1)}$\arabic*}), leftmargin=*]

	\item \textbf{Analyticity:} $M^{(1)}(z)$ is analytic in $\mathbb{C}\setminus\Sigma^{(1)}$, where $
		\Sigma^{(1)}=(\cup_{j=0}^n[E_j,\widehat{E}_j])\cup_{i=1}^4(\Sigma_i\cup\Sigma_i^*)$ (see Figs.~\ref{fig:Hirota71} and~\ref{fig:Hirota72}).
	
	\item \textbf{Jump condition:} $M^{(1)}_+(z)=M^{(1)}_-(z)J_{M}^{(1)}(z)$ for $z\in\Sigma^{(1)}$, where the jump matrix is given by
	\begin{equation}
		J_{M}^{(1)}(z)=\begin{cases}
			-ie^{i(f_0x+(\alpha g_0+\beta h_0)t-(B_j^fx+(\alpha B_j^g+\beta B_j^h)t+\phi_j-\delta_j)/2)\hat{\sigma}_3}\sigma_1,&z\in(E_j,\widehat{E}_j),\\[2.5ex]
			\begin{pmatrix}1&0\\-\overline{r_1(z)}e^{-2it\theta(z)-2\delta(z)}&1\end{pmatrix},&z\in\Sigma_1\cup\Sigma_3,\\[2.5ex]
			\begin{pmatrix}1&r_1(z)e^{2it\theta(z)+2\delta(z)}\\0&1\end{pmatrix},&z\in\Sigma_1^*\cup\Sigma_3^*,\\[2.5ex]
			\begin{pmatrix}1&\dfrac{r_1(z)e^{2it\theta(z)+2\delta(z)}}{1-|r_1(z)|^2}\\0&1\end{pmatrix},&z\in\Sigma_2\cup\Sigma_4,\\[2.5ex]
			\begin{pmatrix}1&0\\ \dfrac{-\overline{r_1(z)}e^{-2it\theta(z)-2\delta(z)}}{1-|r_1(z)|^2}&1\end{pmatrix},&z\in\Sigma_2^*\cup\Sigma_4^*.
		\end{cases}
	\end{equation}
	
	\item \textbf{Normalization at infinity:} As $z\to\infty$,  $M^{(1)}(z)=I+\mathcal{O}(z^{-1})$.
	
	\item \textbf{Local behavior:} For $p\in\{E_j,\widehat{E}_j\}_{j=0}^n$,
	$
		M^{(1)}(z)=\mathcal{O}((z-p)^{-1/4}),\quad z\to p.
	$
\end{enumerate}
\end{RH}

Let
\begin{equation}
	E(z)=
	\begin{cases}
		M^{(1)}(z)M^{(glo)}(z)^{-1}, & z\in\mathbb{C}\setminus \cup_{i=1}^2U_i, \v\\
		M^{(1)}(z)M^{(loc)}(z)^{-1}, & z\in U_i,\quad i=1,2,
	\end{cases}
\end{equation}
where $M^{(glo)}(z)$ and $M^{(loc)}(z)$ are the global and local parametrices for $M^{(1)}(z)$, respectively. The neighborhoods $U_i$ are defined by
\begin{equation}
	U_i=
	\begin{cases}
		\{z:|z-z_i|\leq c_i\}, & \xi\text{ belongs to the Zakharov-Manakov region IV,} \v\\
		\emptyset, & \xi\text{ belongs to the fast-decay region IV,}
	\end{cases}
\end{equation}
with
\begin{equation}
	c_i=\min_{j=0,...,n}\left\{\frac{|\widehat{E}_{j_i}-z_i|}{2},\frac{|E_j-z_i|}{2},\frac{|z_1-z_2|}{2}\right\}.
\end{equation}

As in transition region I, $M^{(glo)}(z)$ satisfies the same global model Riemann--Hilbert problem, whose solution is explicitly given by \eqref{M_glo}. For the local parametrix, it remains only to consider the case in which $\xi$ belongs to the Zakharov--Manakov region.

\begin{RH} $M^{(loc)}(z)$ satisfying the following Riemann--Hilbert problem:

\begin{enumerate}[label=(\textbf{$M^{(loc)}$\arabic*}), leftmargin=*]

	\item \textbf{Analyticity:} $M^{(loc)}(z)$ is analytic in $U_i\setminus\Sigma^{(1)}$.
	
	\item \textbf{Jump condition:} $M^{(loc)}_+(z)=M^{(loc)}_-(z)J_{M}^{(1)}(z)$ for $z\in\Sigma^{(1)}\cap U_i$.
	
	\item \textbf{Matching condition:} As $t\to\infty$, $M^{(loc)}(z)$ matches $M^{(glo)}(z)$ on the boundary $\partial U_i$ of $U_i$.
\end{enumerate}
\end{RH}

To construct the solution, we need a suitable conformal map that captures the local behavior of the phase function near the saddle point $z_i$. We have the following local expansion:
\begin{equation}
	\theta(z)=\theta(z_i)-\theta^{(z_i,2)}(\xi)(z-z_i)^2+\mathcal{O}\left((z-z_i)^3\right),\quad z\to z_i,
\end{equation}
where
\begin{equation}\label{theta_2}
	\theta^{(z_i,2)}(\xi):=\frac{\partial_zF(z_i;\xi)}{w(z_i)}>0.
\end{equation}

Let
\begin{equation}
	\zeta=2(z-z_i)\sqrt{\theta^{(z_i,2)}(\xi)}t^{1/2}.
\end{equation}

This local Riemann--Hilbert problem can be solved explicitly using the parabolic-cylinder parametrix given in \cite{Fan2026,Its1981}. To this end, we obtain
\begin{equation}
	M^{(loc)}(z)=I+\frac{
		t^{-1/2}}{2(z-z_i)\sqrt{\theta^{(z_i,2)}(\xi)}}\begin{pmatrix}
		0 & \beta_{21} \\
		\beta_{12} & 0
	\end{pmatrix}+\mathcal{O}(t^{-1}),\quad t\to\infty,
\end{equation}
where
\begin{equation}\label{beta}
	\beta_{12}=\frac{\sqrt{2\pi}e^{\frac{1}{4}(\pi i-\log(1-|r_0|^2))}}{r_0\Gamma(i\log(1-|r_0|^2)/2\pi)},\quad\beta_{21}=\frac{\log(1-|r_0|^2)}{2\pi\beta_{12}},
\end{equation}
\begin{equation}
	\begin{aligned}
		r_{0}& =-\overline{r_1(z_i)}\left(2\sqrt{\theta^{(z_i,2)}(\xi)}\right)^{\frac{i}{2\pi}\log(1-|r_1(z_i)|^2)} \\
		& \qquad \times \exp\left\{\frac{w(z_i)}{2\pi i}\left[\sum_{j=0}^n\delta_j\int_{E_j}^{\widehat{E}_j}\frac{i\mathrm{d}s}{w_+(s)(s-z_i)}\right]
+\frac{\log(1-|r_1(z_i)|^2)\log(c_i)}{2\pi i}-2it\theta(z_i)\right\} \\
		&\qquad \times \exp\left\{\frac{w(z_i)}{2\pi i}\int_{(-\infty,z_i)\setminus(\cup_{j=0}^n(E_j,\widehat{E}_j))}\left[\frac{\chi_{(z_i-c_i,z_i)}\log(1-|r_1(z_i)|^2)}
{w(z_i)(s-z_i)}-\frac{\log(1-|r_1(s)|^2)}{w(s)(s-z_i)}\right]\mathrm{d}s\right\},
	\end{aligned}
\end{equation}
and $\chi_I$ denotes the characteristic function of the interval $I$.

\begin{figure}[!t]
	\centering
	\includegraphics[scale=0.33]{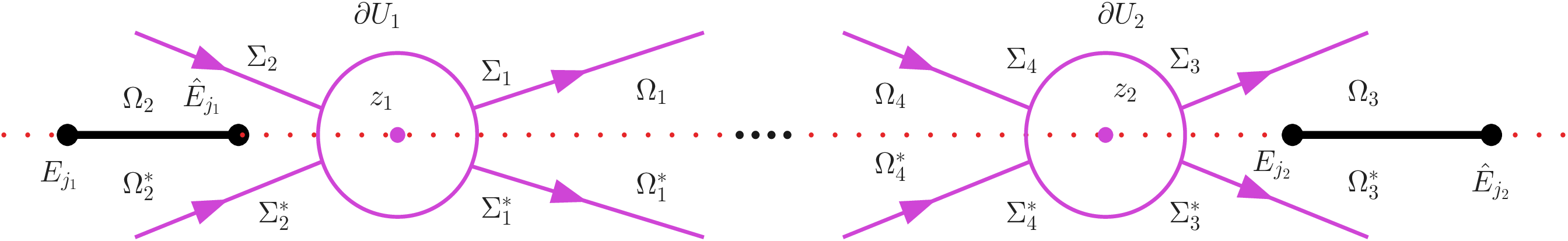}
	\caption{The contours $\Sigma^{(E)}$.}
	\label{fig:Hirota8}
\end{figure}
	
\begin{RH} $E(z)$ satisfies the following Riemann--Hilbert problem:

\begin{enumerate}[label=(\textbf{$E$\arabic*}), leftmargin=*]
	
\item \textbf{Analyticity:} $E(z)$ is analytic in $\mathbb{C}\setminus \Sigma^{(E)}$, where $
		\Sigma^{(E)}=\cup_{i=1}^4(\Sigma_i\cup\Sigma_i^*)\cup\partial U_1\cup\partial U_2\setminus \cup_{i=1}^2U_i$ (see  Fig.~\ref{fig:Hirota8}).
		
	\item \textbf{Jump condition:} $E_+(z)=E_-(z)J_{E}(z)$, where the jump matrix is given by
	\begin{equation}
		J_{E}(z)=\left\{
		\begin{array}{ll}
			M^{(glo)}(z)J_M^{(1)}(z)M^{(glo)}(z)^{-1}, & z\in\Sigma^{(E)}\setminus\cup_{i=1}^2\partial U_i, \v\\
			M^{(loc)}(z)M^{(glo)}(z)^{-1}, & z\in\partial U_i,\quad i=1,2.
		\end{array}\right.
	\end{equation}
	
	\item \textbf{Asymptotics at infinity:} As $z\to\infty$, $E(z)=I+\mathcal{O}(z^{-1})$.
\end{enumerate}
\end{RH}

By the standard small-norm Riemann--Hilbert theory, there exists a unique solution to the Riemann--Hilbert problem for $E$ for sufficiently large positive $t$. As $\xi$ belongs to the Zakharov--Manakov region  III, we have
\begin{equation}\label{E_2}
	E(z)=I+\frac{E_1}{z}+\mathcal{O}(z^{-2}),\quad z\to\infty,
\end{equation}
where
\begin{equation}\label{E_3}
	E_1=\sum_{i=1}^2\frac{t^{-1/2}}{2\sqrt{\theta^{(z_i,2)}(\xi)}}M^{(glo)}(z_i)
	\begin{pmatrix}
		0 & \beta_{21} \\
		\beta_{12} & 0
	\end{pmatrix}M^{(glo)}(z_i)^{-1}+\mathcal{O}(t^{-1}),\quad t\to+\infty.
\end{equation}
As $\xi$ belongs to the fast-decay region IV, we have
\begin{equation}\label{E_24}
	E(z)=I+\mathcal{O}(t^{-1}).
\end{equation}

\section{Proof of Theorem ~\ref{theom}}\label{Proof-Regions}

We now establish the uniform asymptotic expansions for the solution $q(x,t)$ by inverting the sequence of transformations performed in the previous sections. For the asymptotics of $q(x,t)$ in transition region I, the transformations $M \mapsto M^{(1)} \mapsto E$ yield, as $t \to +\infty$,
\begin{equation}
	M(z)=e^{-\delta(\infty)\sigma_3}E(z)M^{(glo)}(z)e^{\delta(z)\sigma_3}G_M(z)^{-1},\quad z\in\mathbb{C}\setminus U_1.
\end{equation}
Using $E(z)=I+\frac{E_1}{z}+\mathcal{O}(z^{-2})$, $M^{(glo)}(z)=I+\frac{M_1^{(glo)}}{z}+\mathcal{O}(z^{-2})$, and $\delta(z)=\delta(\infty)+\frac{\delta^{(1)}}{z}+\mathcal{O}(z^{-2})$ as $z\to\infty$, we obtain
\begin{equation}
	M(z)=I+\frac{1}{z}\left(e^{-\delta(\infty)\hat{\sigma}_3}(E_1+M_1^{(glo)})+\delta^{(1)}\sigma_3\right)+\mathcal{O}(z^{-2}),\quad z\to\infty.
\end{equation}

By the reconstruction formula \eqref{inversion formula}, together with \eqref{M_glo_12} and \eqref{E_1}, we obtain
\begin{equation}
	\begin{aligned}
		q\left(x,t\right) & =2i\lim_{z\to\infty}zM_{12}=2ie^{-2\delta(\infty)}\left((E_1)_{12}+\left(M_1^{(glo)}\right)_{12}\right) \\
		& =e^{-2\delta(\infty)}q^{(AG)}(x,t;\boldsymbol{E},\boldsymbol{\widehat{E}},\boldsymbol{\phi}-\boldsymbol{\delta})
+\frac{2e^{-2\delta(\infty)}}{|(\theta^{(3,\hat{j}_1)}(\widehat{\xi}_{j_1}))^{2/3}|}\nu(s)(H_1)_{11}(\widehat{E}_{j_1})^2t^{-1/3}+\mathcal{O}(t^{-\varrho}).
	\end{aligned}
\end{equation}

A direct calculation gives
\begin{equation}\label{H_I}
	\begin{aligned}
		H_{\widehat{E}_{j_1}} & =\frac{2e^{-2\delta(\infty)}}{|(\theta^{(3,\hat{j}_1)}(\widehat{\xi}_{j_1}))^{2/3}|}(H_1)_{11}(\widehat{E}_{j_1})^2 \\
		& =\frac{e^{i(t\theta(\widehat{E}_{j_1};\xi)-\phi_{j_1}+\delta_{j_1})-2\delta(\infty)}}{|\theta^{(3,\hat{j_1})}(\widehat{\xi}_{j_1})|^{1/3}}\left(\frac{\prod_{j=0}^n(\widehat{E}_{j_1}-E_j)}{\prod_{j=0,\cdots,n,j\neq j_1}(\widehat{E}_{j_1}-\widehat{E}_j)}\right)^{1/2} \\
		&\qquad  \times\left(\frac{\Theta(\mathcal{A}(\widehat{E}_{j_1})+\boldsymbol{C}(x,t;\boldsymbol{\phi}-\boldsymbol{\delta})+\mathcal{A}(\mathcal{D})+\boldsymbol{K})\Theta(\mathcal{A}(\infty)+\mathcal{A}(\mathcal{D})+\boldsymbol{K})}{\Theta(\mathcal{A}(\widehat{E}_{j_1})+\mathcal{A}(\mathcal{D})+\boldsymbol{K})\Theta(\mathcal{A}(\infty)+\boldsymbol{C}(x,t;\boldsymbol{\phi}-\boldsymbol{\delta})+\mathcal{A}(\mathcal{D})+\boldsymbol{K})}\right)^2.
	\end{aligned}
\end{equation}

Similarly, in transition region II, we have
\begin{equation}
	q(x,t)=e^{-2\delta(\infty)}q^{(AG)}(x,t;\boldsymbol{E},\boldsymbol{\widehat{E}},\boldsymbol{\phi}-\boldsymbol{\delta})-\tilde{H}_{E_{j_2}}\nu(s)t^{-1/3}+\mathcal{O}(t^{-\varrho}),
\end{equation}
where
\begin{equation}\label{H_II}
	\begin{aligned}
		\tilde{H}_{E_{j_2}} & =\frac{e^{\pi i/2+i(t\theta(E_{j_2})-\phi_{j_2}+\delta_{j_2})-2\delta(\infty)}}{|\theta^{(3,j_2)}(\xi_{j_2})|^{1/3}}\left(\frac{\prod_{j=0}^n(E_{j_2}-\widehat{E}_j)}{\prod_{j=0,\cdots,n,j\neq j_2}(E_{j_2}-E_j)}\right)^{1/2} \\
		& \qquad \times\left(\frac{\Theta(\mathcal{A}(E_{j_2})+\boldsymbol{C}(x,t;\boldsymbol{\phi}-\boldsymbol{\delta})+\mathcal{A}(\mathcal{D})+\boldsymbol{K})\Theta(\mathcal{A}(\infty)+\mathcal{A}(\mathcal{D})+\boldsymbol{K})}{\Theta(\mathcal{A}(E_{j_2})+\mathcal{A}(\mathcal{D})+\boldsymbol{K})\Theta(\mathcal{A}(\infty)+\boldsymbol{C}(x,t;\boldsymbol{\phi}-\boldsymbol{\delta})+\mathcal{A}(\mathcal{D})+\boldsymbol{K})}\right)^2.
	\end{aligned}
\end{equation}

For the asymptotics of $q(x,t)$ in the Zakharov--Manakov region III, using \eqref{E_3}, we obtain
\begin{equation}
	\begin{aligned}
		q\left(x,t\right) & =2i\lim_{z\to\infty}zM_{12}=2ie^{-2\delta(\infty)}\left((E_1)_{12}+\left(M_1^{(glo)}\right)_{12}\right) \\
		& =e^{-2\delta(\infty)}q^{(AG)}(x,t;\boldsymbol{E},\boldsymbol{\widehat{E}},\boldsymbol{\phi}-\boldsymbol{\delta}) \\
		& \qquad +\sum_{j=1}^2\frac{ie^{-2\delta(\infty)}}{\sqrt{\theta^{(z_j,2)}(\xi)}}\left(\beta_{21}M_{11}^{(glo)}(z_j)^2-\beta_{12}M_{12}^{(glo)}(z_j)^2\right)t^{-1/2}+\mathcal{O}(t^{-1}).
	\end{aligned}
\end{equation}

Finally, for the asymptotics of $q(x,t)$ in the fast-decay region IV, using \eqref{E_24}, we have
\begin{equation}
	\begin{aligned}
		q\left(x,t\right) & =2i\lim_{z\to\infty}zM_{12}=2ie^{-2\delta(\infty)}\left((E_1)_{12}+\left(M_1^{(glo)}\right)_{12}\right) \\
		& =e^{-2\delta(\infty)}q^{(AG)}(x,t;\boldsymbol{E},\boldsymbol{\widehat{E}},\boldsymbol{\phi}-\boldsymbol{\delta})+\mathcal{O}(t^{-1}).
	\end{aligned}
\end{equation}
This completes the proof of Theorem ~\ref{theom}.

%\addcontentsline{toc}{section}{Acknowledgements}	

\v \v\noindent {\bf Acknowledgements}. This work was supported by the National Natural Science Foundation of China
(No. 12471242), and Beijing Natural Science Foundation (No. 1262023).

\v\noindent {\bf Data Availability Statements}. The data that supports the findings of this work are available within the
paper.

\v\noindent {\bf Conflict of Interest}. The authors have no conflicts to disclose.

\end{document}